\documentstyle[aps,preprint,amssymb,subeqn,epsfig,eqsecnum,floats,cite]{revtex}
\tighten
\def \be{\begin{equation}}
\def \ee{\end{equation}}
\def \bea{\begin{eqnarray}}
\def \eea{\end{eqnarray}}
\def \ben{\begin{enumerate}}
\def \een{\end{enumerate}}
\def \bit{\begin{itemize}}
\def \eit{\end{itemize}}
\def \heff{H_{\mathrm{eff}}}
\def \hadron{{\mathrm{had}}}
\def \Oi{{\mathcal O}}
\def \branch{{\cal B}}

\def \Im{{\text{Im}}\,}
\def \Re{{\text{Re}}\,}
\def \GeV{{\text{GeV}}}

\def \av#1{\left\langle #1\right\rangle}
\def \B{\bar{B}}
\def \bm{\boldmath}
\def \braket#1#2#3{\langle #1|#2| #3\rangle}
\def \cl#1{{#1\%\ \mathrm{C.L.}}}
\def \cp{\mathrm{CP}}
\def \cpt{\mathrm{CPT}}
\def \dis{\displaystyle}
\def \ea{{\it et al.}}
\def \eq#1{Eq.~(\ref{#1})}

\def \fig#1{Fig.~\ref{#1}}

\def \nnu{\nonumber}
\def \ol#1{\overline{#1}}
\def \rf{Ref.~\cite}

\def \sh{\hat{s}}

\def \Sec#1{Sec.~\ref{#1}}
\def \sm{\mathrm{SM}}

\def \rcontcc{R_{\mathrm {cont}}^{c\bar{c}}}

\def \rrespsi{R_{\mathrm {res}}^{J/\psi}}
\def \cseff {c_7^{\text{eff}}}
\def \ceff {c_9^{\text{eff}}}

\def \ceffstar{c_9^{\text{eff}*}}
\def \eff{\mathrm{eff}}
\def \fb{\mathrm{FB}}
\def \high{\mathrm{high}}
\def \low{\mathrm{low}}
\def \nr{\mathrm{nr}}
\def \a{\alpha}
\def \b{\beta}

\def \g{\gamma}
\def \G{\Gamma}
\def \d{\delta}
\def \epsi{\epsilon}

\def \k{\kappa}
\def \l{\lambda}

\def \m{\mu}
\def \n{\nu}

\def \p{\pi}

\def \s{\sigma}

\def \t{\tau}
\def\ann#1#2#3{{\it Annu. Rev. Nucl. Part. Sci.\/} {\bf#1} (19#2) #3}
\def\euro#1#2#3{{\it Eur. Phys. J.\/} {\bf C#1} (19#2) #3}

\def\ibid#1#2#3{\emph{ibid.} {\bf#1} (19#2) #3}

\def\ib#1#2#3{{\bf#1} (19#2) #3}

\def\np#1#2#3{{\it Nucl.~Phys.\/}~{\bf B#1} (19#2) #3}

\def\pl#1#2#3{{\it Phys.~Lett.\/}~{\bf B#1} (19#2) #3}

\def\prd#1#2#3{{\it Phys.~Rev.\/}~{\bf D#1} (19#2) #3}
\def\prdn#1#2#3{{\it Phys.~Rev.\/}~{\bf D#1} (20#2) #3}
\def\prl#1#2#3{{\it Phys.~Rev.~Lett.\/}~{\bf #1} (19#2) #3}

\def\rmp#1#2#3{{\it Rev. Mod. Phys.\/} {\bf #1} (19#2) #3}
\def\zpc#1#2#3{{\it Z.~Phys.\/}~{\bf C#1} (19#2) #3}
\begin{document}
\preprint{\setlength{\baselineskip}{1.5em}
\vbox{\vspace{-1cm}
\hbox{FISIST/09/2000/CFIF}
\hbox{SISSA/76/2000/EP}
\hbox{hep-ph/0008210}
\hbox{August 2000}}}
\draft
\title{\Large\bf Looking for Novel \bm$\cp$-Violating Effects in 
\bm$\B\to K^* l^+l^-$}
\author{\sc 
F. Kr\"uger$^{1,}$\thanks{E-mail address: krueger@gtae3.ist.utl.pt} 
and E.~Lunghi$^{2,}$\thanks{E-mail address: lunghi@sissa.it}}
\address{$^1$Centro de F\'{\i}sica das Interac\c{c}\~{o}es Fundamentais (CFIF),
Departamento de F\'{\i}sica,  \\ 
Instituto Superior T\'ecnico,  Av. Rovisco Pais,  1049-001 Lisboa, Portugal}
\address{$^2$SISSA-ISAS, Via Beirut 2-4, 34013 Trieste, Italy\\
and INFN,  Sezione di Trieste, Trieste, Italy}
%
%
\maketitle
\begin{abstract}
The $\cp$-violating asymmetries in the exclusive 
decay $\B\to K^*l^+l^-$ $(l=e,\m,\t)$ are predicted to be exceedingly 
small in the standard model (SM), thereby offering an opportunity to 
assess various new-physics scenarios. We derive quantitative predictions 
for various integrated 
observables in $\B\to K^*\mu^+\mu^-$ decay in the presence of
physics beyond the SM with additional $\cp$ phases and an  
extended operator basis. In particular, a model-independent analysis of 
$\cp$ asymmetries that require the presence of 
unitarity phases, in addition to $\cp$ violation, is performed.
We find that in the 
low dimuon invariant mass region $2m_{\m}\leqslant M_{\mu^+\mu^-} < M_{J/\psi}$, 
the $\cp$ asymmetries are highly suppressed by small dynamical phases,
assuming that new physics is unlikely to 
significantly alter the Wilson coefficients of the operators governing
two-body hadronic $B$ decays. Taking into account current experimental 
data on the measured $b\to s\g$ rate and the upper limit on $\branch(B^0\to K^{*0} \m^+\m^-)$, 
$\cp$-violating effects of a few per cent are estimated,
even in the presence of new physics with $\cp$ phases of $O(1)$.
By contrast, in the high dimuon invariant mass region 
$M_{\psi'}< M_{\mu^+\mu^-}\leqslant (M_B-M_{K^*})$ significant 
$\cp$-violating effects  
are possible.~Given a branching ratio of 
$1.8\times 10^{-6}$, the $\cp$ asymmetries can be quite substantial 
($\sim 20 \%$ or more), and thus may serve as a means of 
discovering physics transcending the SM.     
\end{abstract}
%
%
\section{Introduction}\label{intro}
$\cp$ violation has been observed so far only in the neutral kaon system.
Within the standard model (SM), the experimental results on 
indirect ($\epsi_K$) and 
direct ($\epsi'/\epsi_K$) $\cp$ violation
can be explained by the 
complex phase of the Cabib\-bo-Ko\-ba\-ya\-shi-Maskawa (CKM) matrix if one 
takes into account the large theoretical uncertainties associated with the 
hadronic matrix elements that enter the analysis of 
$\epsi'/\epsi_K$  \cite{epsilonprime}. 
It therefore remains an open question whether the 
CKM mechanism of $\cp$ violation can account for the new experimental result 
on $\epsi'/\epsi_K$.

A great deal of effort is given to the study of $\cp$ violation in the $B$ 
system, which will provide invaluable information on the pattern of 
$\cp$ violation and open up the possibility to look for new physics.
In this paper we are concerned with the exclusive decay
$\B\to K^* l^+l^-$ which is of special interest 
because (i) it probes the 
underlying effective Hamiltonian describing 
flavour-changing neutral current (FCNC) processes in $B$ decay; (ii) 
the analysis of $\cp$-violating effects in decays governed by 
$b\to s l^+l^-$ may offer a deeper insight into the 
mechanism of $\cp$ violation since the SM prediction for $\cp$ asymmetries
is extremely small, typically $\lesssim 10^{-3}$ \cite{cp-phases:btos}; 
and (iii) 
the process $\B\to K^* \m^+\m^-$ is a very promising decay 
mode since it is
likely to be observed in the next round of 
$B$ physics experiments. The most stringent limit has been set by 
CDF of $\branch(B^0\to K^{*0} \m^+\m^-) < 4.0\times 10^{-6}$ 
at the $\cl{90}$ \cite{limits:exc:exp} to be compared with the 
SM prediction of $(1.9\pm 0.7)\times10^{-6}$ \cite{ali:etal}.

The object of the present work is to explore the possibility of 
sizable $\cp$ asymmetries in $\B\to K^*l^+l^-$ decay, 
whose observation would clearly indicate the presence of physics beyond the 
SM. 
We perform a largely model-independent analysis 
by considering a new-physics scenario with additional $\cp$ phases and an 
extended operator basis, taking into 
account existing experimental data on $\B\to X_s \g$ and the upper bound on  
$B^0\to K^{*0}\m^+\m^-$.

Our paper is organized as follows.~Section \ref{effective:hamiltonian}
contains the SM operator basis and short-distance 
matrix element for the process $\B\to K^* l^+ l^-$. A formalism for 
dealing with real $c\bar{c}$ intermediate states such as 
$J/\psi,\psi'$ entering via the decay chain 
$\B\to K^* V_{c\bar{c}}\to K^* l^+l^-$, is described.
In \Sec{eff:ham:np}, we briefly discuss possible extensions of the SM
including those with an extended operator basis.
Section \ref{diff:decay-rate} is concerned with the parametrization 
of the hadronic matrix elements and gives the differential decay 
spectrum as well as the forward-backward asymmetry.~The corresponding 
$\cp$-violating observables are discussed in 
\Sec{sec:cp:observables}.~Numerical estimates for integrated ob\-ser\-va\-bles 
in $\B\to K^* \m^+\m^-$ decay in the presence of new 
physics are given in \Sec{results:num}. Particular attention is paid to
the $\cp$-violating partial-rate asymmetry and the $\cp$-violating 
effect related to the angular distribution of $\m^-$ in $B$ and $\B$ 
decays. Our conclusions are contained in  \Sec{conclusions}.

\section{Effective Hamiltonian}\label{effective:hamiltonian}
\subsection{Short-distance contributions}
The effective Hamiltonian for the decay $\B \rightarrow K^*l^+l^-$
in the standard model (SM) is given by \cite{bmm,review}
\be\label{heff}
\heff = -\frac{4G_F}{\sqrt{2}}V_{tb}^{}V_{ts}^{*}\left\{
\sum\limits_{i = 1}^{10} c_i (\m)\Oi_i(\m) + \l_u\{c_1(\m)[\Oi^u_1(\m)-\Oi_1^{}(\m)]+c_2(\m)[\Oi_2^u(\m)-\Oi_2^{}(\m)]\}\right\},
\ee
where we have used the unitarity of the 
CKM matrix,
$\l_u\equiv V_{ub}^{}V^*_{us}/V_{tb}^{}V^*_{ts}$, and the 
operator basis is defined as follows:
\bea\label{operatorbasis}
{\Oi_1}&=& (\bar{s}_{ \alpha} \gamma_\mu P_L c_{ \beta})
(\bar{c}_{ \beta} \gamma^\mu P_Lb_{ \alpha}),\nnu\\
{\Oi_1^u}&=& (\bar{s}_{ \alpha} \gamma_\mu P_L u_{ \beta})
(\bar{u}_{ \beta} \gamma^\mu P_Lb_{ \alpha}),\nnu\\
{\Oi}_2 &=& (\bar{s}_{ \alpha} \gamma_\mu P_L c_{ \alpha})
(\bar{c}_{ \beta} \gamma^\mu P_L b_{ \beta}),\nnu\\
{\Oi}_2^u &=& (\bar{s}_{ \alpha} \gamma_\mu P_L u_{ \alpha})
(\bar{u}_{ \beta} \gamma^\mu P_L b_{ \beta}),\nnu\\
{\Oi}_3 &=& (\bar{s}_{ \alpha} \gamma_\mu P_L b_{ \alpha})\sum_{q=u,d,s,c,b}
(\bar{q}_{ \beta} \gamma^\mu P_L q_{ \beta}),\nnu\\
{\Oi}_4&=& (\bar{s}_{ \alpha} \gamma_\mu P_L b_{ \beta})
\sum_{q=u,d,s,c,b}(\bar{q}_{ \beta} \gamma^\mu P_L q_{ \alpha}),\nnu\\
{\Oi}_5&=& (\bar{s}_{ \alpha} \gamma_\mu P_L b_{ \alpha})
\sum_{q=u,d,s,c,b}(\bar{q}_{ \beta} \gamma^\mu P_R q_{ \beta}),\nnu\\
{\Oi}_6&=& (\bar{s}_{ \alpha} \gamma_\mu P_L b_{ \beta})
\sum_{q=u,d,s,c,b}(\bar{q}_{ \beta} \gamma^\mu P_R q_{ \alpha}),\nnu\\ 
{\Oi}_7&=&\frac{e}{16 \pi^2}\bar{s}_{\alpha} \sigma_{\mu \nu} (m_b P_R + m_s P_L) b_{\alpha}F^{\mu \nu},\nnu\\
{\Oi}_8& =& \frac{g_s}{16 \pi^2}\bar{s}_{\alpha} 
T_{\alpha \beta}^a \sigma_{\mu \nu} (m_b P_R + m_s P_L)
b_{\beta} G^{a \mu \nu},\nnu\\
{\Oi}_9&=& \frac{e^2}{16 \pi^2} \bar{s}_\alpha \gamma^{\mu} P_L b_\alpha
\bar{l} \gamma_{\mu} l,\nnu\\
{\Oi}_{10}&=&\frac{e^2}{16 \pi^2} \bar{s}_\alpha \gamma^{\mu} P_L
b_\alpha \bar{l} \gamma_{\mu}\gamma_5 l,
\eea
where $\a$, $\b$ are colour indices, $a$ labels the SU(3) generators, 
and $P_{L,R}= (1\mp \g_5)/2$. 
Evolution of the Wilson coefficients $c_i(\m)$ in \eq{heff} 
from the weak scale $\m=M_W$ down to the low energy scale 
$\m=m_b$ by means of  the renormalization group equations (RGE's)
then leads to the QCD-corrected matrix element
in next-to-leading logarithmic approximation \cite{bmm,review}
\bea\label{mael:short}
\lefteqn{{\cal M}= \frac{G_F \a}{\sqrt{2} \p}V_{tb}^{}V_{ts}^{\ast}
\Bigg\{\Bigg[(\ceff-c_{10}) \braket{K^*(k)}{\bar{s}\g_{\m}P_L b}{\B(p)}}\nnu\\& & \mbox{}- \frac{2\cseff}{s}\braket{K^*(k)}{\bar{s}i \s_{\m\n}q^{\n}\left(m_b P_R+ m_s P_L\right) b}{\B(p)}\Bigg]\bar{l}\g^{\m} P_L l + 
(c_{10}\to -c_{10}) \bar{l}\g^{\m} P_R l\Bigg\}. 
\eea
Here $q=p-k$, $s\equiv q^2$ is the invariant mass of the 
lepton pair, and the effective Wilson coefficient $\ceff$ has the form
\be\label{c9eff:def}
c_9^{\eff}= c_9 + Y(s),
\ee
with 
\bea\label{wilson:c9}
Y(s)&=& g(m_c,s)(3 c_1 + c_ 2 +3c_3 + c_4 + 3c_5 + c_6)
+ \l_u[g(m_c,s)- g(m_u,s)](3 c_1 + c_ 2)\nnu\\ 
&&\mbox{}-\frac{1}{2} g(m_s,s)(c_3 + 3 c_4)
-\frac{1}{2}g(m_b,s)(4c_3 + 4c_4 +3c_5+c_6)\nnu\\
&&\mbox{}+\frac{2}{9}(3c_3 +c_4+3c_5+c_6)+ \cdots,
\eea
where the ellipsis represents the order $\a_s$ correction 
to the matrix element 
of the operator $\Oi_9$, which can be regarded as a contribution to the 
form factors \cite{lw}, and hence will be omitted in the calculations  
that follow.~Table \ref{table:SD} summarizes our results for the Wilson 
coefficients $c_i(m_b)$. Observe that $\cseff$, $c_9$, and $c_{10}$ are real 
in the framework of the SM. The function $g(m_i, s)$ in the above 
formula arises from the one-loop 
contributions of the four-quark operators $\Oi_1$--$\Oi_6$, 
and is given by (at $\m=m_b$)
\bea\label{loopfunc}
\lefteqn{g(m_i,s)=-\frac{8}{9}\ln(m_i/m_b)+\frac{8}{27}+\frac{4}{9}y_i
-\frac{2}{9}(2+y_i)\sqrt{|1-y_i|}}\nnu\\[.7ex]
&&\times\left\{
\Theta(1-y_i)\left[\ln\left(\frac{1 + \sqrt{1-y_i}}{1 - \sqrt{1-y_i}}\right)-i\p\right]+ \Theta(y_i-1) 2\arctan\frac{1}{\sqrt{y_i-1}}\right\},
\eea
with $y_i = 4m_i^2/s$. 
%
%
\begin{table}
\caption{Numerical values of the Wilson coefficients $c_1,\dots, c_{10}$ at $\m=m_b$ within the SM.}\label{table:SD}
\begin{tabular}{cccccccccc}
$c_1$& $c_2$&$c_3$& $c_4$& $c_5$ & $c_6$& $\cseff$& $c_9$&$c_{10}$\\
\hline
$-0.249$& $+1.108$&$+0.011$& $-0.026$& $+0.007$ & $-0.031$& $-0.314$& $+4.216$ & $-4.582$
\end{tabular}
\end{table}
%
%
This expression reduces to
\be
g(0,s)=\frac{8}{27}- \frac{4}{9}\ln(s/m_b^2) + \frac{4}{9} i \pi,
\ee
in the limit $m_i\to 0$. A few remarks are in order here. 

(i) The Wilson coefficient $\ceff$, \eq{c9eff:def}, 
has absorptive parts for $s> 4m_u^2 $ and $s> 4 m_c^2$, 
and thus contains dynamical (unitarity) phases. As will become clear, 
these are prerequisites, besides a $\cp$-violating phase,
for observing $\cp$ asymmetries in partial rates.
Since the remaining coefficients $\cseff$ and $c_{10}$ do not contain
any strong phases, the unitarity phases below the $c\bar{c}$ 
threshold are generated by light quark contributions, 
whereas for $s> 4m_c^2$ they arise mainly from $c\bar{c}$ intermediate 
states. 

(ii) Within the SM, $\cp$ violation in $b\to s l^+l^-$ transition 
is caused by the ratio of CKM factors appearing in $\ceff$, 
namely $\l_u \sim\l^2$ ($\l\equiv V_{us}\simeq 0.22$), 
which  is further reduced by a 
factor of order $(3c_1+c_2)/c_9\simeq0.085$. As a result, 
the effective Hamiltonian for $b \to sl^+l^-$
essentially involves only one independent CKM factor $V_{tb}^{}V_{ts}^*$,
so that $\cp$ violation in this channel is unobservably small. 
Thus, the numerical effect of $\l_u$ in \eq{wilson:c9} is 
negligible for decays based on the transition $b\to s l^+l^-$.

\subsection{Resonance contributions}
In addition to the short-distance contributions discussed so far, 
there are possible quark antiquark resonant intermediate 
states like $\phi, J/\psi, \psi'$ etc.
Since the $s$-quark contributions in \eq{wilson:c9} are 
suppressed by small Wilson coefficients, and terms proportional to 
$\l_u$ may be dropped in the case of $b\to s$ transition, 
we are left with the $J/\psi$ family.

Following the procedure in \rf{fklms:res}, we implement the 
charmonium resonances utilizing $e^+e^-$ annihilation data.   
The absorptive part of the one-loop function 
$g$, \eq{loopfunc}, 
can be related to the experimentally accessible quantity 
\be
R(s)\equiv \s_{\text{tot}}(e^+e^-\to {\text{hadrons}})/
\s(e^+e^-\to \m^+\m^-)
\ee
by virtue of the optical theorem. Specifically, it is found that
\cite{fklms:res}
\be\label{g:absorptive}
\Im g(m_c, s)=\frac{\p}{3}R^{J/\psi}(s),
\ee
with $R^{J/\psi}(s)\equiv\rcontcc (s)+ \rrespsi (s)$,
whereas the dispersive part $\Re g(m_c,s)$ may be obtained via a
once-subtracted dispersion relation
\be\label{dispersion}
g(m_c,s) = g(m_c,0) + \frac{s}{3}\int_{4M_{\p}^2}^\infty
\frac{R^{J/\psi}(s')}{s'(s'-s-i\epsi)}d s', \quad \epsi\to +0,
\ee
with $g(m_c,0)= -8/9\ln(m_c/m_b)-4/9$. 
The continuum contributions, $\rcontcc$, can be determined using 
experimental data from \rf{res:data:burk},
while the narrow resonances are well described by a relativistic 
Breit-Wigner distribution \cite{res:data:burk,res:data:jeger}
\be\label{breit-wigner}
\rrespsi (s)=\sum_{V=J/\psi, \psi',\dots}
\frac{9s}{\a^2}\frac{\branch(V\to l^+l^-)\G^V_{\text{tot}}\G^V_{\hadron}}{(s-M_V^2)^2+ M_V^2\G^{V^2}_{\text{tot}}},
\ee
with the properties of the vector mesons summarized in \rf{pdg}.

\begin{figure}[ht]
\begin{center}
\epsfig{file=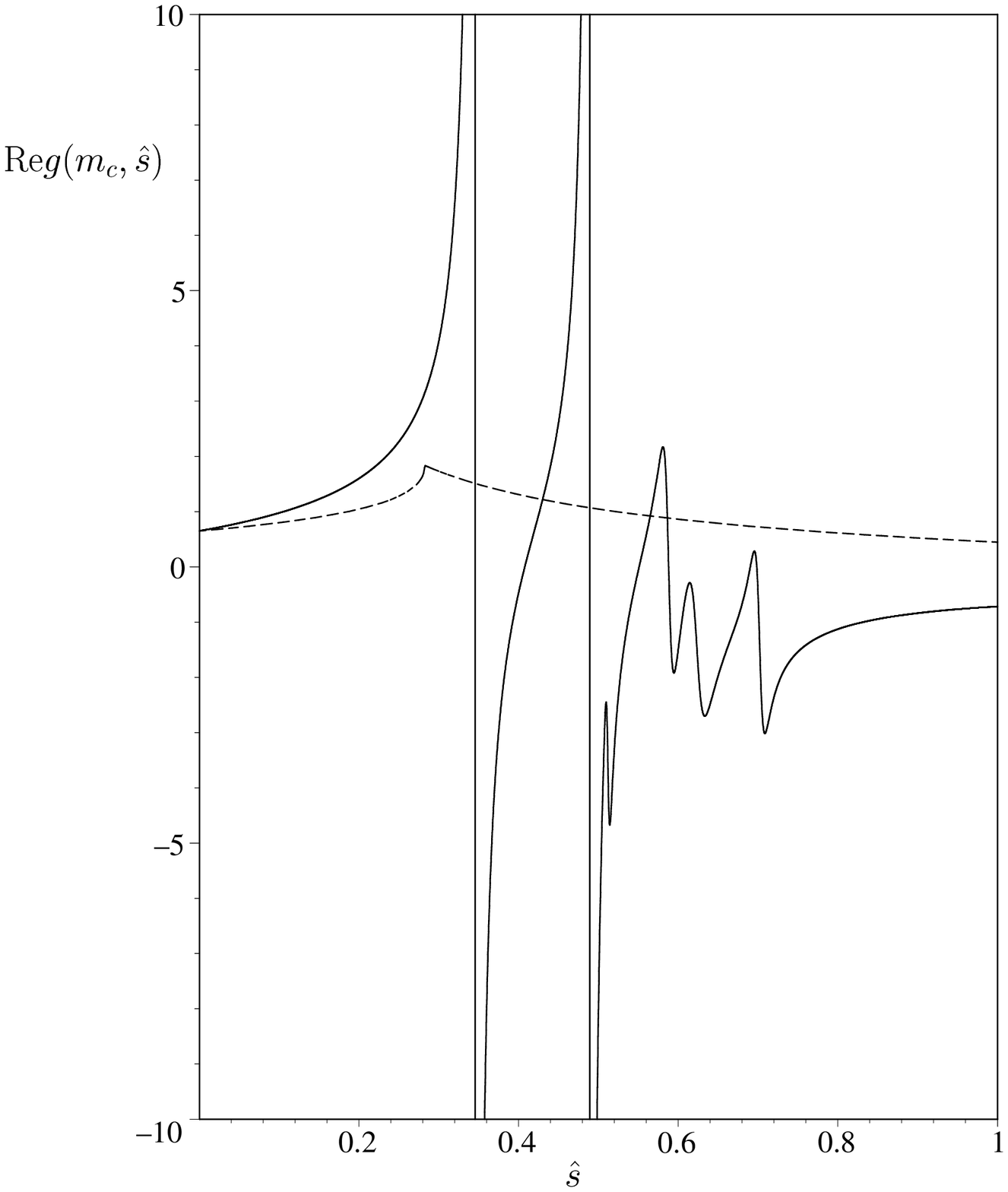,height=3.5in}\hspace{1em}
\epsfig{file=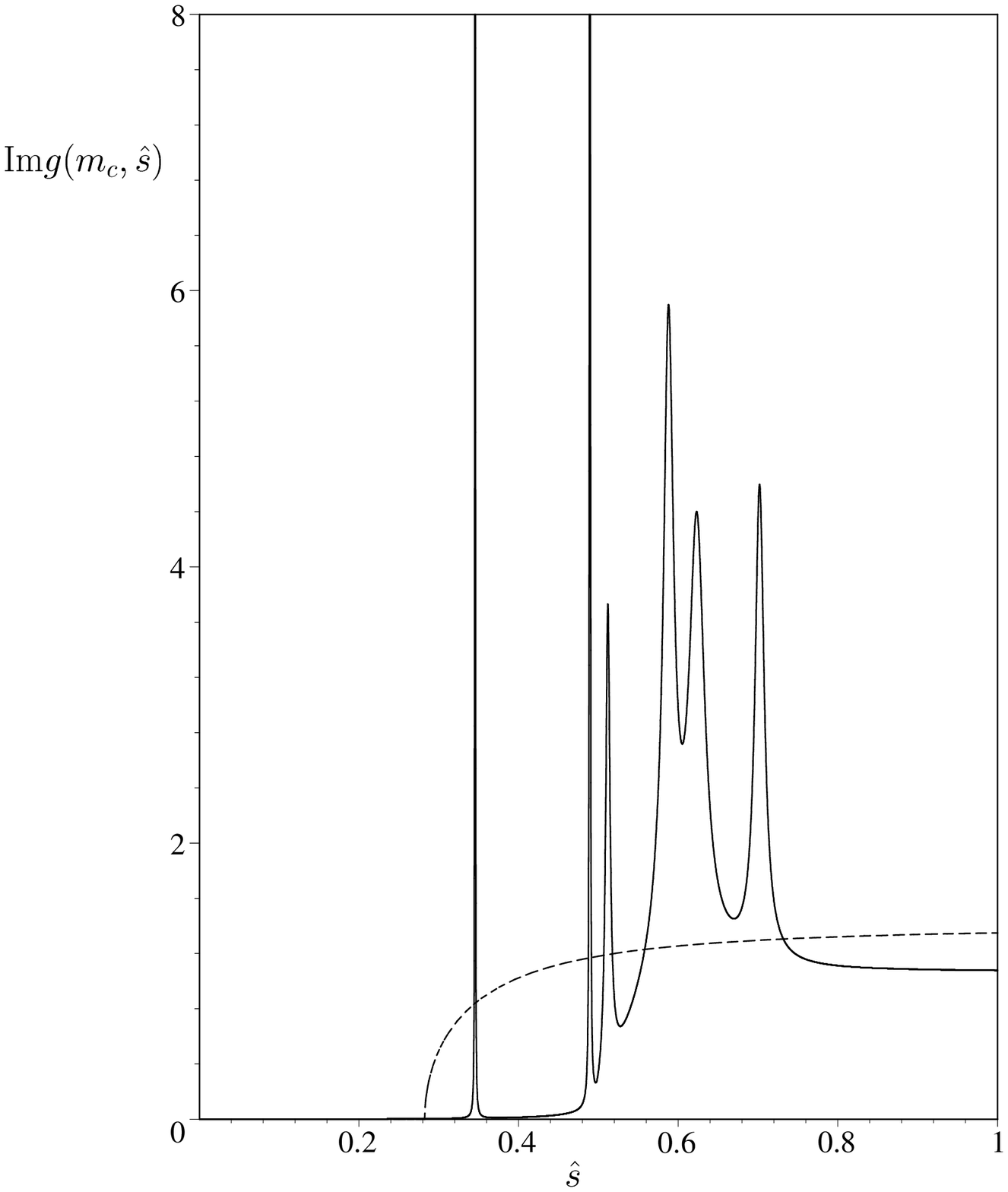,height=3.5in}\vspace{0.5em}
\caption[]{The predicted $\sh$ dependence of $g(m_c,\sh)$ 
using experimental data on $e^+e^-\to \mathrm{hadrons}$ 
via a dispersion relation, \eq{dispersion}, 
including the effects of $c\bar{c}$ resonances (solid curve).
Note that $\Im g(m_c, \sh)\neq 0$ is entirely due to unitarity phases.
Also shown is the result of the perturbative calculation  
according to \eq{loopfunc} (dashed curve).
\label{comp:per:non-pert}}
\end{center}
\end{figure}

To account for experimental data on direct 
$J/\psi$ production via the relation 
\be\label{rel:factorize}
\branch(\B\to K^* V_{c\bar{c}}\to K^* l^+l^-)=
\branch(\B\to K^* V_{c\bar{c}})\branch(V_{c\bar{c}}\to l^+l^-),
\ee 
where $V_{c\bar{c}}=J/\psi, \psi',\dots$, one usually modifies 
the Breit-Wigner distribution in \eq{breit-wigner} by introducing 
an ad hoc factor
$\k_V$ \cite{lw,kappa-factor}.~This suggests that the factorization ansatz
inherent in the approaches that have been advocated to incorporate 
resonance effects in $b\to s l^+l^-$ (see, e.g., \rf{ali:hiller}) 
is inadequate for two-body hadronic decays of $B$ mesons.
Using the set of form factors that will be discussed below, one finds
$\k_{J/\psi}=1.7$, $\k_{\psi'}=2.4$, whereas for the remaining resonances
we shall take  (as in \rf{ali:etal}) $\k=2$. In \fig{comp:per:non-pert}, we 
show the real and imaginary parts of $g(m_c,\sh)$ based  
on \eq{dispersion}, as a function of $\sh= s/M_B^2$.
%
%
\section{Effective Hamiltonian and new physics}\label{eff:ham:np}
The new-physics contributions to the decay mode 
$b\to s l^+l^-$ can manifest itself 
in two distinct ways: (i) The absolute values and phases of the 
Wilson coefficients at the electroweak scale are modified. (ii) 
New operators in addition to the ones defined in \eq{operatorbasis} arise.  
We consider them in turn.

\subsection{Wilson coefficients and new physics}
The non-standard contributions can be divided into three main groups 
\cite{np:review}: First, 
models with tree-level contributions to the 
four-quark operators such as supersymmetry (SUSY) 
with broken R-parity or models with $Z$-mediated flavour-changing 
neutral currents.
Second, models with contributions at one-loop level with new particles 
running in the loop like SUSY with conserved R-parity or models with  
four quark generations. 
Third, models with no significant effect on the 
above Wilson coefficients including multi-Higgs-doublet models with natural 
flavour conservation and left-right symmetric models. 

Let us begin with the Wilson coefficients $c_3,\dots,c_6$ of the 
QCD penguin operators. It follows from the RGE analysis 
that their numerical values at $\m=m_b$ are essentially 
determined by the Wilson coefficient $c_2$ of the four-quark 
operator ${\Oi}_2$ at the electroweak scale. 
Thus, in order to have considerable deviations from the SM predictions for 
the coefficients $c_3$--$c_6$, one needs large   
new-physics contributions to the short-distance coefficient $c_2$, 
which in turn would affect the theoretical branching ratio for 
two-body non-leptonic $B$ decays.
Recent studies \cite{ali:kramer:lue} suggest, however, 
that the short-distance coefficients of 
the SM can account for existing data if one allows departures
from the naive factorization prescription.~

For the numerical analysis here, we adopt the SM values for the 
Wilson coefficients $c_1, \dots, c_6$ summarized in Table \ref{table:SD}. 
Furthermore, it is convenient for later use to parametrize the new-physics 
contributions to the remaining coefficients $\cseff$, $\ceff$, 
and $c_{10}$ by the following ratios (defined at the scale $\m=m_b$): 
\begin{mathletters}\label{np:wilson}
\be\label{r7}
R_7= c_7^{\eff}/c_7^{\eff, \sm}\equiv |R_7|e^{i \phi_7}, 
\ee
\be
R_9 = c_9/c_9^{\sm}\equiv |R_9|e^{i \phi_9},
\ee
\be
R_{10} = c_{10}/c_{10}^{\sm}\equiv |R_{10}|e^{i \phi_{10}},
\ee
\end{mathletters}%
where we recall \eq{c9eff:def}, and
\be\label{coeff:sm:np}
c_i=c_i^{\sm}+ c_i^{\mathrm{New}},
\ee
with $c_i^{\sm}$ shown in Table \ref{table:SD}. 
Note that the Wilson coefficient $c_{10}$ in the above does not depend on the 
renormalization scale, and thus $c_{10}\equiv c_{10}(M_W)$.

\subsection{Extended operator basis}
It is conceivable that physics beyond the SM induces new operator structures 
containing scalar, pseudoscalar, and tensor interactions, 
in addition to the SM operator basis, \eq{operatorbasis}. 

The Wilson coefficients of the 
scalar operators
\bea\label{operatorbasis:scalar}
\Oi^S_9 &=& \bar s_\alpha P_L b_\alpha \bar l  l , \nnu\\
\Oi^S_{10} &=& \bar s_\alpha  P_L b_\alpha \bar l \gamma_5 l,\nnu\\ 
\Oi^{S'}_9 &=& \bar s_\alpha P_R b_\alpha \bar l  l,\nnu\\ 
\Oi^{S'}_{10} &=& \bar s_\alpha  P_R b_\alpha \bar l \gamma_5 l,
\eea
involve the lepton mass in most extensions of the SM, and hence 
will give only small contributions in the case of $l=e$ or $\m$.
Moreover, possible tensor-type operators  
$\bar{s} \s_{\m \n} b \bar{l}\s^{\m \n} l$ and
$\bar{s} i \s_{\m \n} b \bar{l}\s_{\a \b} l \epsilon^{\m \n \a\b}$
can be safely neglected since their numerical contributions have been found 
to be small \cite{int:tensor}. 

This leaves an extended operator basis which consists of the 
SM operators and the opposite-chirality operators
\cite{int:tensor,left-right}
\bea\label{operatorbasis:chirality}
\Oi_7'&=&\frac{e}{16 \pi^2}\bar{s}_{\alpha} \sigma_{\mu \nu} (m_b P_L + m_s P_R) b_{\alpha}F^{\mu \nu},\nnu\\
\Oi_8' &=& \frac{g_s}{16 \pi^2}\bar{s}_{\alpha} 
T_{\alpha \beta}^a \sigma_{\mu \nu} (m_b P_L + m_s P_R)
b_{\beta} G^{a \mu \nu},\nnu\\
\Oi_9'&=& \frac{e^2}{16 \pi^2} \bar{s}_\alpha \gamma^{\mu} P_R b_\alpha
\bar{l} \gamma_{\mu} l,\nnu\\
\Oi_{10}'&=&\frac{e^2}{16 \pi^2} \bar{s}_\alpha \gamma^{\mu} P_R
b_\alpha \bar{l} \gamma_{\mu}\gamma_5 l.
\eea

\section{Decay distribution for \bm$\B\to K^*\lowercase{l^+l^-}$}\label{diff:decay-rate}
\subsection{Form factors}\label{formfactors}
The hadronic matrix elements appearing in 
\eq{mael:short} can be expressed in terms of $s$-dependent form factors 
(recall $s\equiv q^2$); namely \cite{stech:wirbel}
\bea\label{ff:btokstar1}
\lefteqn{\braket{{K^*}(k)}{\bar{s}\g_{\m}P_{L,R} b}{\bar{B}(p)} = i\epsi_{\m\n\a\b} \epsi^{\n*}p^{\a}q^{\b} \frac{V(s)}{M_B+ M_{{K^*}}} \mp\frac{1}{2}\Bigg\{\epsi_{\m}^*(M_B+ M_{{K^*}})A_1(s)}\nnu \\[.7ex]
&-& (\epsi^*\cdot q) (2p-q)_{\m}\frac{A_2(s)}{M_B+M_{{K^*}}} -\frac{2M_{{K^*}}}{s}(\epsi^*\cdot q)\big[A_3(s) - A_0(s)\big]q_{\m}\Bigg\}\ , 
\eea
with the convention $\epsi_{0123} = +1$, $q=p-k$, and 
\bea\label{ff:btokstar3}
\lefteqn{\braket{{K^*}(k)}{\bar{s}i \s_{\m\n}q^{\n}P_{R,L} b}{\bar{B}(p)}
= - i\epsi_{\m\n\a\b} \epsi^{\n*}p^{\a}q^{\b}T_1(s)
\pm\frac{1}{2}\Bigg\{[\epsi_{\m}^*(M_B^2-M_{{K^*}}^2)}\nnu\\[.7ex]
& -& (\epsi^*\cdot q)(2p - q)_{\m}] T_2(s)
+(\epsi^*\cdot q)\Bigg[q_{\m} - \frac{s}{M_B^2-M_{{K^*}}^2}(2p-q)_{\m}\Bigg]T_3(s)\Bigg\},
\eea
where $T_1(0)=T_2(0)$, and $\epsi^{\m}$ is the 
$K^*$ polarization vector. The form factor $A_3$ can be written in terms 
of $A_1$ and $A_2$, i.e. 
\bea\label{ff:btokstar2}
A_3(s) = \frac{M_B+M_{{K^*}}}{2M_{{K^*}}} A_1(s) -\frac{M_B-M_{{K^*}}}{2M_{{K^*}}} A_2(s),
\eea   
with the relation $A_3(0)=A_0(0)$. The terms proportional to $q_{\m}$ in 
Eqs.~(\ref{ff:btokstar1}) and (\ref{ff:btokstar3}) do not contribute 
to the differential decay rate when the final leptons are 
massless.
For the results presented below, we adopt the $B\to K^*$ form 
factors of \rf{ali:etal} which have been obtained using light cone sum rule 
(LCSR) results, and are displayed in Table \ref{table:formfactors}.
Throughout our discussion, we assume the above form factors to be real, 
in the absence of final-state interactions.
%
%
\begin{table}
\caption{LCSR predictions for the $B\to K^*$ form factors with $f(\sh) = f(0)\exp(c_1 \sh+ c_2 \sh^2)$, $\sh=s/M_B^2$ \cite{ali:etal}. Recall that 
$A_3$ is given in terms of $A_1$ and $A_2$ via \eq{ff:btokstar2}.}\label{table:formfactors}
\begin{tabular}{cccccccc}
 & $V$&$A_1$ &$A_2$ &$A_0$ &$T_1$ &$T_2$ &$T_3$ \\ \hline
$f(0)$ &$0.457$ &$0.337$ &$0.282$ &$0.471$ &$0.379$ &$0.379$ &$0.260$\\
$c_1$ &$1.482$ &$0.602$ &$1.172$ &$1.505$ &$1.519$ &$0.517$ &$1.129$\\
$c_2$ &$1.015$ &$0.258$ &$0.567$ &$0.710$ &$1.030$ &$0.426$ &$1.128$\\
\end{tabular}
\end{table}
%
%
\subsection{Differential decay spectrum}
Squaring the matrix element (\ref{mael:short}), summing over spins, 
and introducing the shorthand notation 
\be\label{def:triangle}
\l (a,b,c) = a^2 + b^2 + c^2 - 2 (a b + b c + a c),
\ee
\be
\hat{M}_i= M_i/M_B, \quad \hat{m_i}=m_i/M_B,\quad \sh=s/M_B^2,     
\ee
\be\label{def:X}
X=\frac{1}{2}\l^{1/2}(1, \sh, \hat{M}_{K^*}^2), 
\ee
the spectrum of $\B\to K^*l^+l^-$ decay with respect to $\sh$ and 
$\theta_l$, the angle between $l^-$ and the outgoing hadron
in the dilepton centre-of-mass system, is \cite{corr:lepton}
\be\label{diff:BtoV}
\frac{d\G(\B\to K^* l^+l^-)}{d \sh\, d\cos\theta_l}
=\frac{G_F^2 \a^2M_B^5}{2^9 \p^5}\,|V_{tb}^{}V_{ts}^{\ast}|^2 X
\sqrt{1-\frac{4\hat{m}_l^2}{\sh}}
[A(\sh) + B(\sh) \cos\theta_l + C(\sh) \cos^2\theta_l].
\ee
The quantities $A$, $B$, $C$ are defined as follows:  
\be\label{double:A}
A(\sh)= \frac{2X^2}{\hat{M}_{K^*}^2}\Bigg[\sh \hat{M}_{K^*}^2 
f_1(\sh) 
+ \frac{1}{4}\Bigg(1 + \frac{2 \sh \hat{M}_{K^*}^2}{X^2}\Bigg) f_2(\sh)
+ X^2 f_3(\sh)+ f_4(\sh)\Bigg]+ 2\hat{m}_l^2 I(\sh), 
\ee
\be\label{double:B}
B(\sh) = 8  X\sqrt{1-\frac{4\hat{m}_l^2}{\sh}}\, 
\Re\{c_{10}^*[\ceff \sh A_x A_y - \cseff 
(A_x B_y + A_y B_x)]\},
\ee
\be\label{double:C}
C(\sh) =  \frac{2X^2}{\hat{M}_{K^*}^2}
\left(1-\frac{4\hat{m}_l^2}{\sh}\right)\Bigg[\sh \hat{M}_{K^*}^2 f_1(\sh) 
- \frac{1}{4} f_2(\sh)- X^2 f_3(\sh) - f_4(\sh)\Bigg],
\ee
with the auxiliary functions 
\be
I(\sh)= 4X^2 f_1(\sh) + f_2(\sh) + f_5(\sh),
\ee
\be
f_1(\sh) = (|\ceff|^2 + |c_{10}|^2) A_x^2 + 
\frac{4|\cseff|^2}{\sh^2}B_x^2
-\frac{4\Re(\cseff\ceffstar)}{\sh} {A}_x {B}_x, 
\ee
\be
f_2(\sh) = {f_1(\sh)}_{x\to y}, \quad f_3 (\sh) = {f_1(\sh)}_{x\to z},
\ee
\bea
f_4(\sh) &=& \frac{1}{2}(1- \sh-\hat{M}_{K^*}^2)\nnu\\
&\times&
\Bigg[(|\ceff|^2 + |c_{10}|^2) {A}_y {A}_z + \frac{4|\cseff|^2}{\sh^2} {B}_y{B}_z
-\frac{2\Re(\cseff\ceffstar)}{\sh}(A_y B_z+A_z B_y)\Bigg],
\eea
\bea
f_5(\sh)= |c_{10}|^2\Bigg[&-&8X^2 A_x^2  -3A_y^2
+ \frac{X^2}{\hat{M}_{K^*}^2}\{[2(1+\hat{M}_{K^*}^2)-\sh]A_z + 2A_y\}A_z\nnu\\
&+&\frac{4X^2}{\sh\hat{M}_{K^*}}\{\hat{M}_{K^*}(A_3-A_0)-[(1-\hat{M}_{K^*}^2)A_z +A_y]\}(A_3-A_0)\Bigg].
\eea
In these equations, 
\be
A_x=\frac{V(\sh)}{1+ \hat{M}_{K^*}},
\quad A_y=(1+ \hat{M}_{K^*})A_1(\sh),
\quad A_z= - \frac{A_2(\sh)}{1+ \hat{M}_{K^*}}, 
\ee
\be
B_x= -T_1(\sh)(\hat{m}_b +\hat{m}_s) , 
\quad B_y= -(1-\hat{M}_{K^*}^2)T_2(\sh)(\hat{m}_b -\hat{m}_s), 
\ee
\be
B_z= \Bigg[T_2(\sh) + \frac{\sh}{1-\hat{M}_{K^*}^2}T_3(\sh)\Bigg]
(\hat{m}_b -\hat{m}_s),
\ee
with the form factors listed in Table \ref{table:formfactors}. 

A further observable of interest is the forward-backward (FB) asymmetry of 
$l^-$, defined as 
\be\label{FB}  
A_{\text{FB}}(\sh)=\frac{\dis\int_0^1 d\cos\theta_l
\frac{d\G}{d\sh\, d\cos\theta_l}-\int_{-1}^0 d\cos\theta_l
\frac{d\G}{d\sh\, d\cos\theta_l}}{\dis\int_0^1 d\cos\theta_l
\frac{d\G}{d\sh\, d\cos\theta_l}+\int_{-1}^0 d\cos\theta_l
\frac{d\G}{d\sh\, d\cos\theta_l}},
\ee
and we obtain
\be\label{fb:BtoV}
A_{\text{FB}}(\sh)= 12 X \sqrt{1-\frac{4\hat{m}_l^2}{\sh}}\,
\frac{\Re\{c_{10}^*[\ceff \sh A_x A_y 
- \cseff(A_x B_y + A_y B_x)]\}}{[3A(\sh)+ C(\sh)]}.
\ee
Note that the Wilson coefficients in the above expressions 
are defined through \eq{coeff:sm:np}. The decay distributions in the presence of new physics with additional operators are discussed further in 
\Sec{results:num}. We now proceed to a discussion of $\cp$-violating 
observables in the process  $\B\to K^*l^+l^-$.
  
\section{\bm$\cp$-violating observables}\label{sec:cp:observables}
Suppose the decay amplitude for $\B\to F$ has the general form  
\be\label{matrixelement:itof}
{\cal A}(\B\to F)=e^{i\phi_1}A_1 e^{i\d_1} +e^{i\phi_2}A_2 e^{i\d_2},   
\ee
where  $A_{1,2}$ are real matrix elements, $\d_i$ and $\phi_i$ are 
the strong phases ($\cp$-conserving) and weak phases ($\cp$-violating)
respectively.
Using $\cpt$ invariance, which requires that the total decay rate for particle and antiparticle be equal, the decay amplitude for the conjugate 
process takes the form 
\be\label{matrixelement:baritof}
\ol{\cal A}(B\to \bar{F})
=e^{-i\phi_1}A_1 e^{i\d_1} +e^{-i\phi_2}A_2 e^{i\d_2},
\ee
giving rise to the $\cp$ asymmetry
\be\label{def:CPasym:gen}
A_{\cp}\equiv\frac{|{\cal A}|^2-|\ol{\cal A}|^2}{|{\cal A}|^2
+|\ol{\cal A}|^2}= \frac{-2 r\sin\phi\sin\d}{1 + 2r\cos\phi\cos\d+ r^2},
\ee
with $r=A_2/A_1$, $\phi=\phi_1-\phi_2$, and $\d=\d_1-\d_2$. 
Notice  that in the limit $r \ll 1$ the asymmetry is approximately 
$A_{\cp}\approx -2 r \sin\phi \sin\d$.
Inspection of \eq{def:CPasym:gen} reveals that a non-zero partial-rate 
asymmetry 
requires $\cp$ violation ($\phi\neq 0$) and the presence of dynamical 
phases ($\d\neq 0$), the latter being provided by the one-loop function
$g(m_i, s)$ present in the Wilson coefficient $\ceff$ [\eq{c9eff:def}].
 
To determine the impact of the strong phases on  
the $\cp$ asymmetries, it is useful to separate those contributions 
to the decay amplitude that generate absorptive parts. 
To this end, we choose $\d_1=\phi_2=0$ in \eq{matrixelement:itof} 
and require that $A_2 e^{i\d_2}$ vanishes when $Y(s)\to 0$. 
Moreover, as discussed in \Sec{eff:ham:np}, 
the corrections to the short-distance 
coefficients of the SM multiplying the absorptive parts of the decay amplitude 
are not expected to be large in many extensions of the SM, nor required by 
current data on two-body hadronic $B$ decays \cite{ali:kramer:lue}. 
Consequently, the strong phases 
in $b\to s l^+l^-$ decay are essentially unaffected by possible new 
interactions transcending the SM, so that the non-standard effects on 
$\cp$ asymmetries can be described by the two parameters $r$ and $\phi$.

\begin{figure}
\begin{center}
\epsfig{file=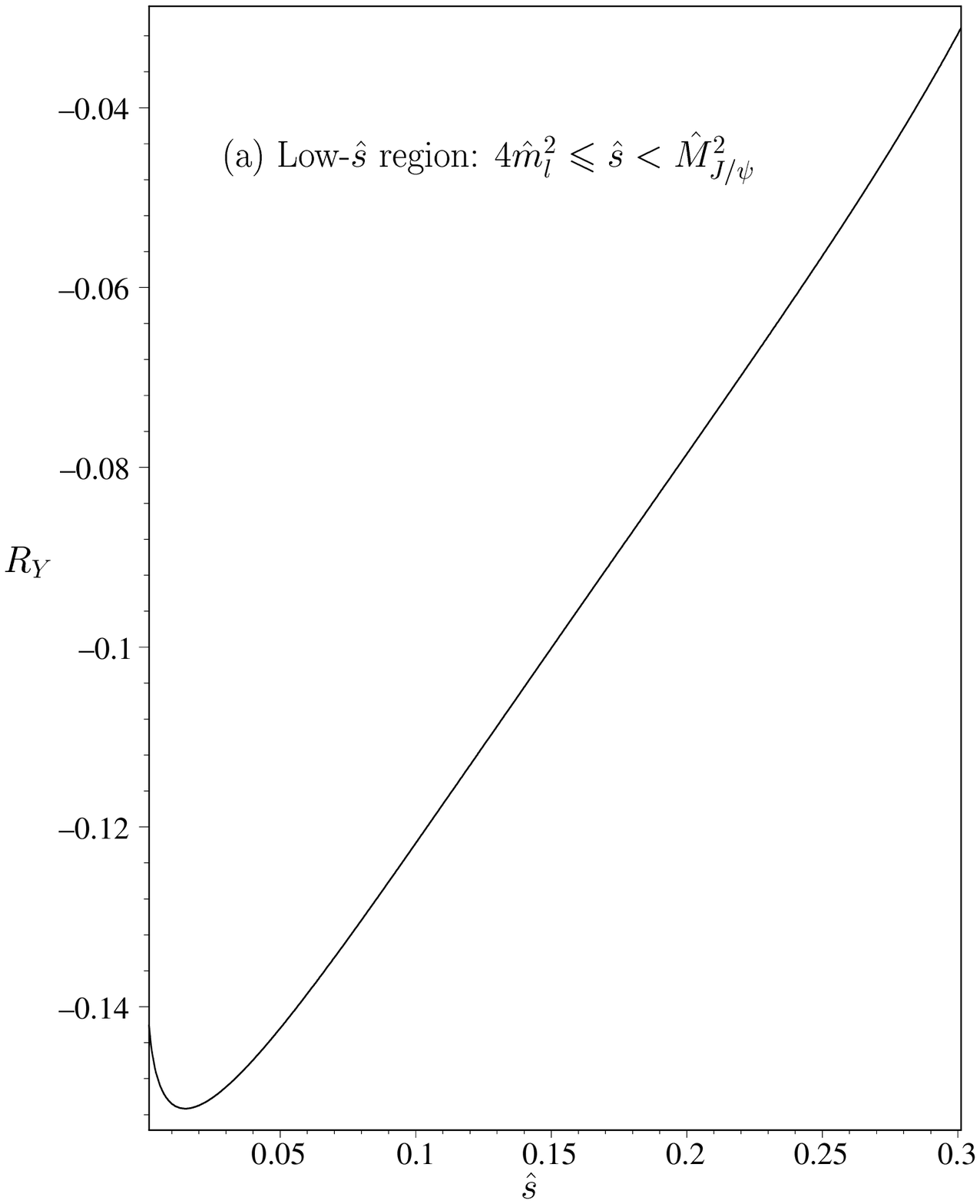,height=3.6in}\hspace{1em}
\epsfig{file=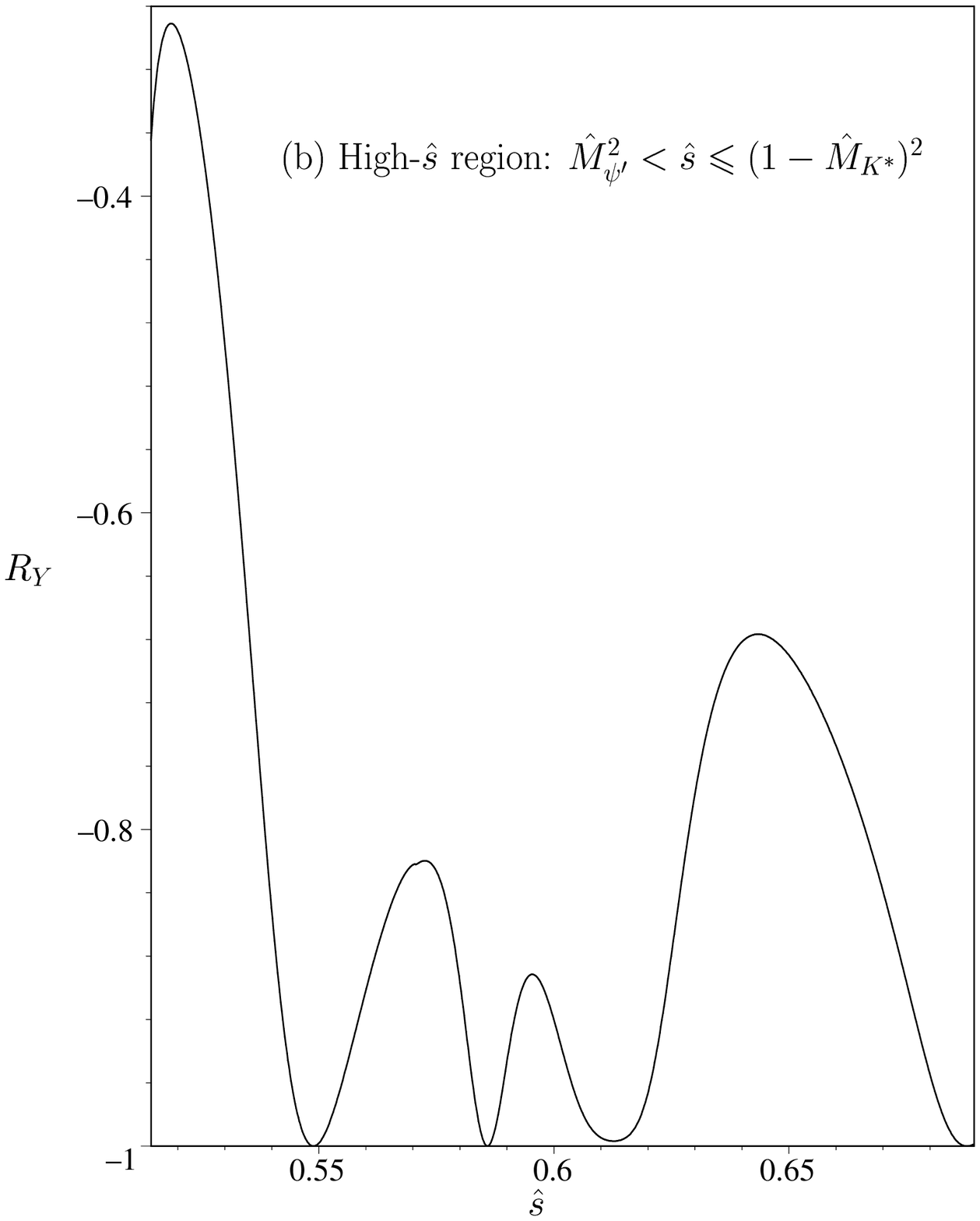,height=3.6in}\vspace{0.5em}
\caption[]{The quantity $R_Y\equiv -\Im[Y(\sh)]/|Y(\sh)|$ in $b\to s l^+l^-$ 
decay vs $\sh$ for the low-$\sh$ (a) 
and high-$\sh$ (b) regions, with $\sh=s/M_B^2$.\label{fig:strong-phase}}
\end{center}
\end{figure}

In order to get a quantitative idea of the magnitude of unitarity phases 
in $b\to s l^+l^-$ transition, we plot in 
\fig{fig:strong-phase} the parameter $R_Y\equiv -\Im[Y(\sh)]/|Y(\sh)|$
for two different regions of the dilepton invariant mass, namely
\be\label{low-high}
4 \hat{m}_l^2\leqslant \sh \leqslant (\hat{M}_{J/\psi}
-\epsi_{\mathrm{cut}})^2, 
\quad (\hat{M}_{\psi'}+\epsilon_{\mathrm{cut}}')^2\leqslant \sh 
\leqslant (1-\hat{M}_{K^*})^2,
\ee
which we refer to as the low-$\sh$ and high-$\sh$ region respectively.\footnote{We use $\epsi_{\mathrm{cut}}=0.2\ \GeV/M_B$ and $\epsi_{\mathrm{cut}}'=0.1\ \GeV/M_B$ in the case of $\m^+\m^-$ in the 
final state.}
%
%
As we alluded to earlier, in the low-$\sh$ region the strong phase is 
suppressed by small Wilson coefficients of the QCD penguin operators,
whereas in the high-$\sh$ region it can be large.

\begin{figure}
\begin{center}
\epsfig{file=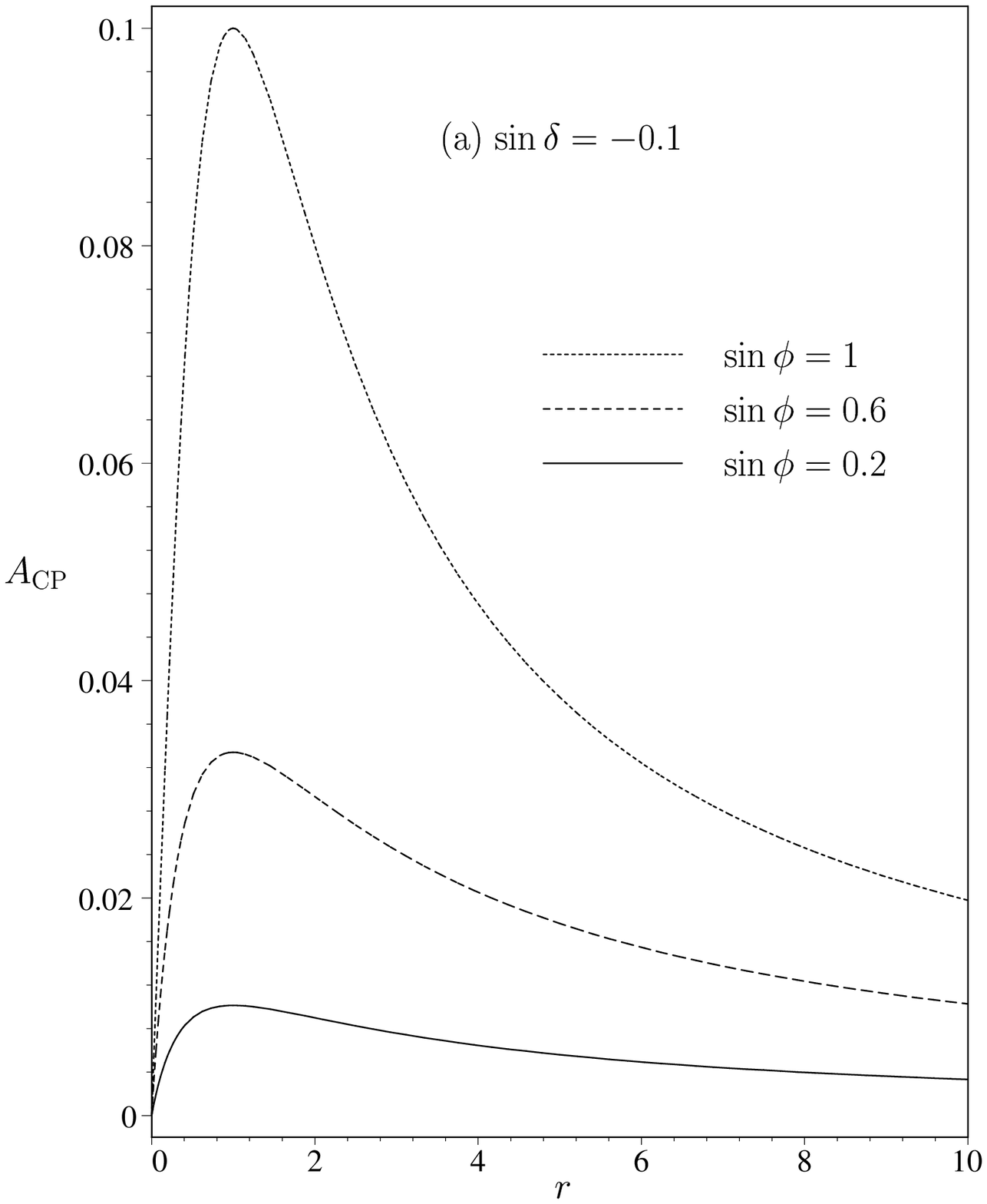,height=3.6in}\hspace{1em}
\epsfig{file=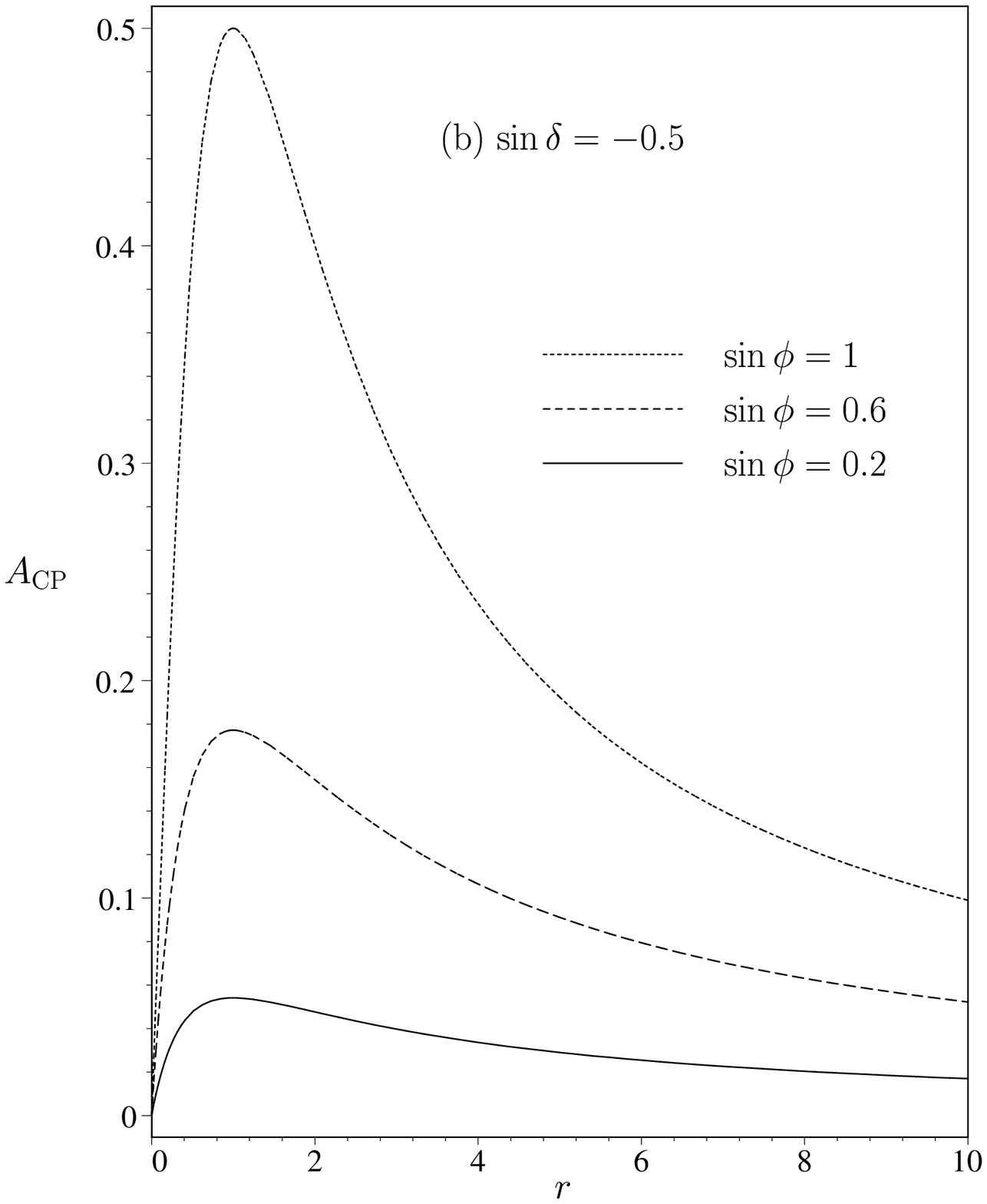,height=3.6in}\vspace{0.5em}
\caption[]{$\cp$-violating asymmetry $A_{\cp}$, \eq{def:CPasym:gen},
as a function of $r$ for (a) the low-$\sh$ region with $\sin \d=-0.1$ and (b) 
the high-$\sh$ region with $\sin\d=-0.5$. \label{fig:cp-violation}}
\end{center}
\end{figure}

Figure \ref{fig:cp-violation} shows the dependence of $A_{\cp}$ 
on $r$ for different choices of the weak phase $\sin\phi$ 
in the low-$\sh$ and high-$\sh$ region.
Observe that the asymmetry is maximized when   
$r=1$ (i.e. the two interfering amplitudes are of comparable size)
but is strongly suppressed if either $r\gg 1$ or $r\ll 1$. 
%
%

Using the two-dimensional decay distribution 
$d\G/d\sh \, d\cos\theta_l$ derived in the preceding section,
we may define the following average $\cp$ asymmetries:   
\be\label{def:CPasymSD}
\av{A_{\cp}^{D}}=\frac{\dis \int_{\sh_0}^{\sh_1} d\sh \int_{D} d\cos\theta_l\frac{d\G_{\text{sum}}}{d \sh\, d\cos\theta_l}}{\dis\int_{\sh_0}^{\sh_1} 
d\sh \int_{-1}^{+1} d\cos\theta_l\frac{d\G_{\text{sum}}}{d \sh\, d\cos\theta_l}},
\quad
\av{A_{\cp}^{S}}=\frac{\dis \int_{\sh_0}^{\sh_1} d\sh \int_{S} d\cos\theta_l\frac{d\G_{\text{diff}}}{d \sh\, d\cos\theta_l}}{\dis\int_{\sh_0}^{\sh_1} 
d\sh \int_{-1}^{+1} d\cos\theta_l\frac{d\G_{\text{sum}}}{d \sh\, d\cos\theta_l}},
\ee
where
\be
\int_{D, S}\equiv \int_0^{1}\mp\int_{-1}^0,
\ee
\be\label{gamma:sum}
\Gamma_{\text{sum}}=\Gamma (\bar{B}\to K^* l^+l^-)+ 
\ol{\G}(B \to\bar{K}^*l^+l^-),
\ee
\be\label{gamma:diff}
\Gamma_{\text{diff}}=\Gamma(\bar{B}\to K^* l^+l^-)- 
\ol{\G}(B \to\bar{K}^*l^+l^-).
\ee
Notice that the asymmetry $A_{\cp}^D$ is a $\cp$-violating effect in 
the angular distribution of $l^-$ (or equivalently the Dalitz-plot 
distribution) in $B$ and $\B$ decays while 
$A_{\cp}^S$ represents the asymmetry in the partial widths of 
these decays.~The special feature of the former is that 
it can be determined even for an untagged equal mixture of $B$ and $\B$ 
events, i.e.~without flavour identification.~We emphasize that both asymmetries are odd under $C\hspace{-0.1em}P$ but 
even under `naive' $T$, and thus vanish in the limit that there are no 
strong phases.\footnote{For a review of $\cp$-odd observables, we refer 
the reader to \rf{valencia}.}

\section{Numerical analysis}\label{results:num}
In the remainder of this paper, we wish to focus on the $\B\to K^* \m^+\m^-$ 
mode. We start this section with a brief discussion of the experimental 
constraints that we shall be using in our subsequent calculations.

\subsection{Experimental constraints}
Constraints on the parameters $R_i$, Eqs.~(\ref{np:wilson}), are 
provided by the following experimental results: (i) The CDF collaboration
has recently obtained the upper bound 
\be\label{cdf:limit}
\branch(B^0\to K^{*0} \m^+\m^-) < 4.0\times 10^{-6}
\ee
at $\cl{90}$ \cite{limits:exc:exp}. 
(ii) The measurement of the 
inclusive branching ratio $\branch(\B\to X_s \g)$ yields \cite{exp:btosg}
\be\label{range:btosg}
2.0\times 10^{-4}< \branch(\B\to X_s \g)<4.5\times 10^{-4}
\quad (\cl{95}).
\ee
Employing the leading-order result for $\branch(\B\to X_s \g)$ (see, e.g.,   
\rf{kagan:neubert:btosg}), useful bounds can be placed on the absolute value of
$R_7$, namely
\be\label{bsg:r7}
0.881 < |R_7| < 1.321.
\ee

\subsection{Integrated observables in \bm$\B \to K^*\m^+\m^-$}
Below we present our predictions (i) for the non-resonant invariant 
mass spectrum of the muon pair, 
and (ii) for the low-$\sh$ and high-$\sh$ region, as defined in 
\eq{low-high}. The resonance effects are taken into 
account by employing Eqs.~(\ref{g:absorptive}) and (\ref{dispersion}). 
\bit
\item The branching ratio resulting from the SM operator basis given in 
\eq{operatorbasis} can be conveniently written as
\bea\label{br:num:gen}
\branch&=&[a_0+ a_1|R_{10}|^2 + a_2 |R_7|^2+ a_3|R_9|^2 +a_4 \Im R_7
+a_5 \Im R_9 +a_6 \Re (R_7 R_9^*)\nnu\\
&&\mbox{}+ a_7 \Re R_7 +a_8\Re R_9]\times  10^{-7}.
\eea
Here we have introduced the coefficients $a_i\equiv \a_i + \b_i$
which allow us to incorporate the effects of new operators ${\Oi}_i'$ 
[\eq{operatorbasis:chirality}] into our computation of the 
branching ratio by simply replacing
\be
(\a_i+\b_i)f(R_i)\to \a_if(R_i+R_i') + \b_i f(R_i-R_i'),
\ee
where we have defined the quantities $R_i'$ analogous to those in 
Eqs.~(\ref{np:wilson}). Our numerical results for the coefficients 
$\a_i$ and $\b_i$ are listed in Table \ref{table:num:res:br-fb}.  

%
%
%
\begin{table}[t]
\caption{Numerical estimate of the coefficients $a_i\equiv \a_i+\b_i$ entering 
the expressions for the integrated branching ratio $\branch$ 
and the average FB asymmetry 
$\av{A_{\fb}}$ in $\B\to K^* \m^+\m^-$ decay, 
as described in the text.\label{table:num:res:br-fb}}
\begin{tabular}{cccc}
$\text{Coefficients}$& $\branch^{\nr}$&$\branch^{\low}$&$\branch^{\high}$\\
\hline
$(\a_0,\b_0)$ &$(0.03,0.12)$ &$(0.02,0.08)$ &$(0.02,0.13)$\\
$(\a_1,\b_1)$ &$(1.89,8.64)$ &$(0.62,3.84)$ &$(0.36,1.76)$\\  
$(\a_2,\b_2)$ &$(1.05,1.06)$ &$(0.95,0.90)$ &$(0.01,0.04)$\\
$(\a_3,\b_3)$ &$(1.62,7.34)$ &$(0.54,3.26)$ &$(0.30,1.50)$\\
$(\a_4,\b_4)$ &$-(0.07,0.18)$ &$-(0.01,0.02)$ &$-(0.02,0.10)$\\
$(\a_5,\b_5)$ &$(0.23,0.93)$ &$(0.01,0.07)$ &$(0.11,0.60)$\\
$(\a_6,\b_6)$ &$-(1.52,3.12)$ &$-(0.90,1.59)$ &$-(0.13,0.52)$\\
$(\a_7,\b_7)$ &$-(0.15,0.30)$ &$-(0.12,0.21)$ &$(0.02,0.07)$\\
$(\a_8,\b_8)$ &$(0.33,1.41)$ &$(0.18,0.91)$ &$-(0.08,0.44)$\\ \hline
$\text{Coefficients}$ &$\av{A_{\fb}}^{\nr}$&
$\av{A_{\fb}}^{\low}$&$\av{A_{\fb}}^{\high}$\\ \hline
$(\a_0,\b_0)$ &$(0.33,0.33)$ &$(0.16,0.16)$ &$-(0.05,0.05)$\\
$(\a_1,\b_1)$ &$-(1.27,1.42)$ &$-(0.69,0.77)$ &$-(0.16,0.18)$\\  
$(\a_2,\b_2)$ &$(3.29,3.29)$ &$(0.94,0.94)$ &$(0.82,0.82)$\\
$(\a_3,\b_3)$ &$(0.25,0.25)$ &$(0.01,0.01)$ &$(0.16,0.16)$
\end{tabular}
\end{table}
%
%
\item  Similarly, we parametrize the average FB asymmetry as 
\be\label{fb:num:gen}
\av{A_{\fb}}=-[\Re R_{10}^*(a_0+ a_1R_7+a_2R_9)+a_3\Im R_{10}] 
\Bigg(\frac{10^{-7}}{\branch}\Bigg),
\ee
with the values of $a_i$ tabulated in Table \ref{table:num:res:br-fb}.
As before, the average FB asymmetry in the presence of chirality-flipped 
operators can be obtained from \eq{fb:num:gen} by the following 
replacements:
\bea
(\a_i+\b_i)f(R_7,R_9,R_{10})&\to&\a_i f(R_7-R_7',R_9-R_9',R_{10}+R_{10}')\nnu\\
&+& \b_i f(R_7+R_7',R_9+R_9',R_{10}-R_{10}').
\eea
\item Our results for the $\cp$ asymmetries are as follows: 
\bea\label{cp:asym:width:low}
\av{A_{\cp}^{S}}^{\low} =
(a_0\Im R_7+a_1\Im R_9)
\Bigg(\frac{10^{-9}}{\hat{\branch}^{\low}}\Bigg), 
\eea
\bea
\av{A_{\cp}^{S}}^{\high} =(a_0\Im R_7+a_1\Im R_9)
\Bigg(\frac{10^{-8}}{\hat{\branch}^{\high}}\Bigg),
\eea
and 
\bea
\av{A_{\cp}^{D}}^{\low} = -a_0 \sin\phi_{10}|R_{10}| 
\Bigg(\frac{10^{-9}}{\hat{\branch}^{\low}}\Bigg),
\eea
\bea\label{cp:asym:angular:high}
\av{A_{\cp}^{D}}^{\high} = -a_0\sin\phi_{10}|R_{10}| 
\Bigg(\frac{10^{-8}}{\hat{\branch}^{\high}}\Bigg),
\eea
$\hat\branch\equiv (\branch+\bar{\branch})/2$ being the $\cp$-averaged 
branching ratio [see Eqs.~(\ref{matrixelement:itof}) and 
(\ref{matrixelement:baritof})]. It is obvious that 
$\av{A_{\cp}^{S}}$ is sensitive to the $\cp$-violating phases 
$\phi_7$ and $\phi_9$, while $\av{A_{\cp}^{S}}$ also probes the phase 
$\phi_{10}$. The predictions for 
the coefficients $a_i$ are reported in Table \ref{table:num:res:cp}. 
\eit
%
%
\begin{table}[t]
\caption{Values of the coefficients $a_i\equiv \a_i+\b_i$ for $\cp$-violating 
observables in  $\B\to K^* \m^+\m^-$.\label{table:num:res:cp}}
\begin{tabular}{ccccc}
\text{Coefficients} & $\av{A_{\cp}^{S}}^{\low}$ &$\av{A_{\cp}^{S}}^{\high}$&
$\av{A_{\cp}^{D}}^{\low}$&$\av{A_{\cp}^{D}}^{\high}$\\
\hline
$(\a_0,\b_0)$ &$-(0.99,1.76)$ &$-(0.25,1.03)$ &$(1.05,1.05)$&$(1.62,1.62)$\\
$(\a_1,\b_1)$ &$(1.20,7.22)$ &$(1.16,5.98)$ &$-$&$-$\\  
\end{tabular}
\end{table}
\subsection{Numerical results and predictions}
We first analyse the $\cp$ asymmetries  in the context of the 
SM operator basis, \eq{operatorbasis}.
To determine the implications of new physics for the $\cp$ asymmetries, 
we adopt the following procedure. The absolute value of $R_7$ is 
chosen such that it satisfies the constraints implied by the 
measured $b\to s \g$ rate, \eq{bsg:r7}, while the remaining parameters 
$R_9$ and $R_{10}$ are required to be 
consistent with the current experimental upper limit on the non-resonant 
branching ratio $\branch(B^0\to K^{*0} \m^+\m^-)$ given in \eq{cdf:limit}.
To gain predictivity we take $|R_7|$ to be unity, so that we end up 
with a set of free parameters comprising $\phi_7$, $\phi_9$, and $|R_9|$. 
In fact, imposing the requirement that the non-resonant branching ratio 
coincides with the experimental upper bound, or, alternatively, 
with the SM prediction, the quantity $R_{10}$ is computed for any given set 
of the parameters $\phi_7, \phi_9, |R_9|$.\footnote{As far as new $\cp$-violating phases are concerned, we note that in the 
context of a specific model one has 
also to take into account the severe constraints on the electric dipole 
moments of electron and  neutron.}
This enables us to consider a scenario where the predicted 
$b\to  s \g$ fraction coincides with the SM expectation
while the non-resonant branching ratio of $\B \to K^* \m^+\m^-$
may well be accessible in the next round of $B$ experiments.

As mentioned above, the value of the strong phase entering the 
amplitude of the $\B\to K^* \mu^+\mu^-$ process depends
on the Wilson coefficients $c_1,\dots, c_6$ which we have assumed to be 
unaffected by new-physics contributions. Hence, the value of the 
strong phase $\sin\d$ is fixed, and estimated to be 
$-0.07$ and $-0.51$ in the low-$\sh$ 
and high-$\sh$ domain respectively. 

Lastly, using the parametrization for the amplitude 
given in \eq{matrixelement:itof}, and employing the integrated expressions
above, we determine $r$ and $\sin\phi$ numerically as a function  of 
$\phi_7,\phi_9$, and $|R_9|$.

\subsubsection{CP asymmetry in the partial widths}
\begin{figure}[p]
\begin{center}
\epsfig{file=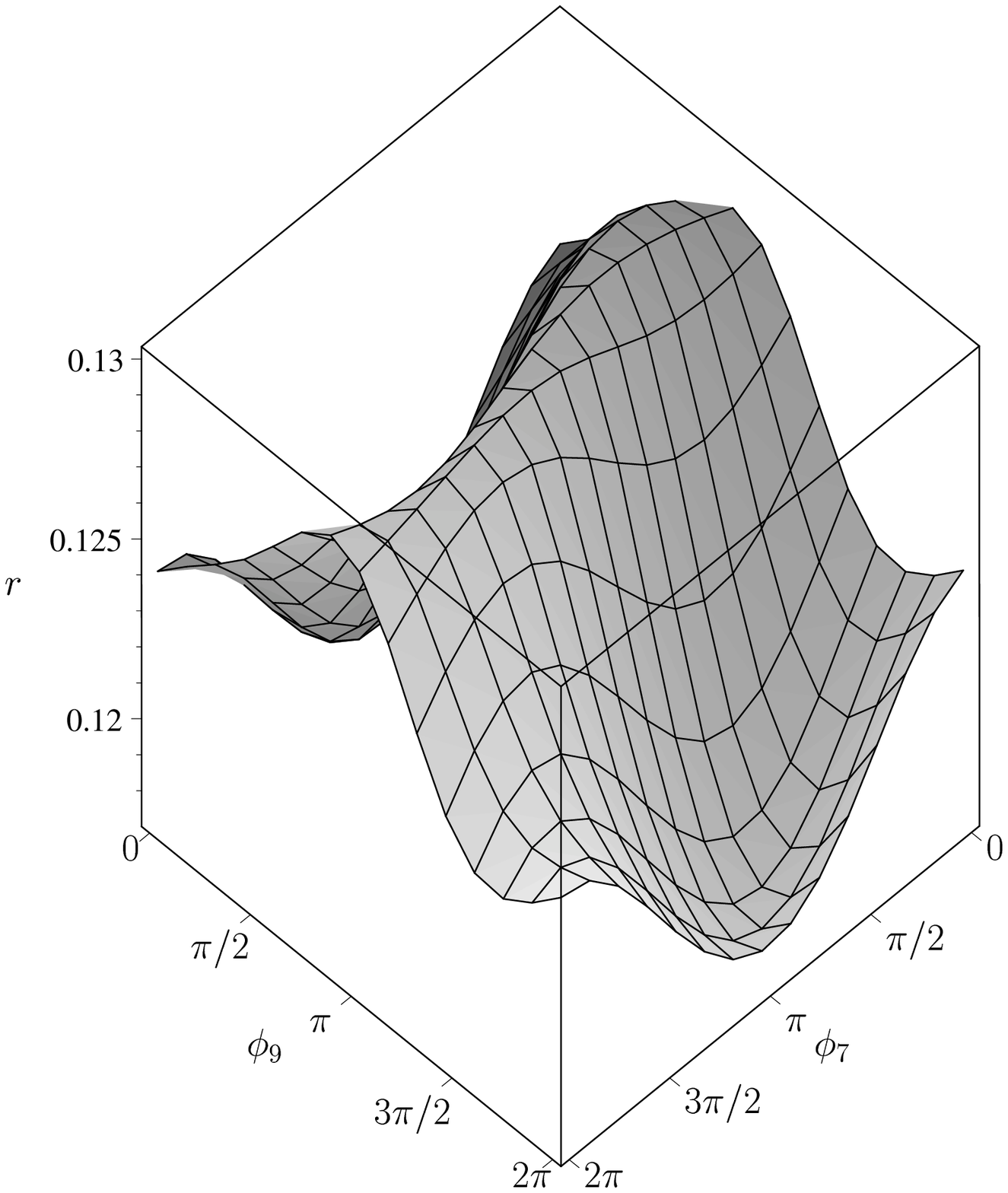,height=3.4in}\hspace{1em}
\epsfig{file=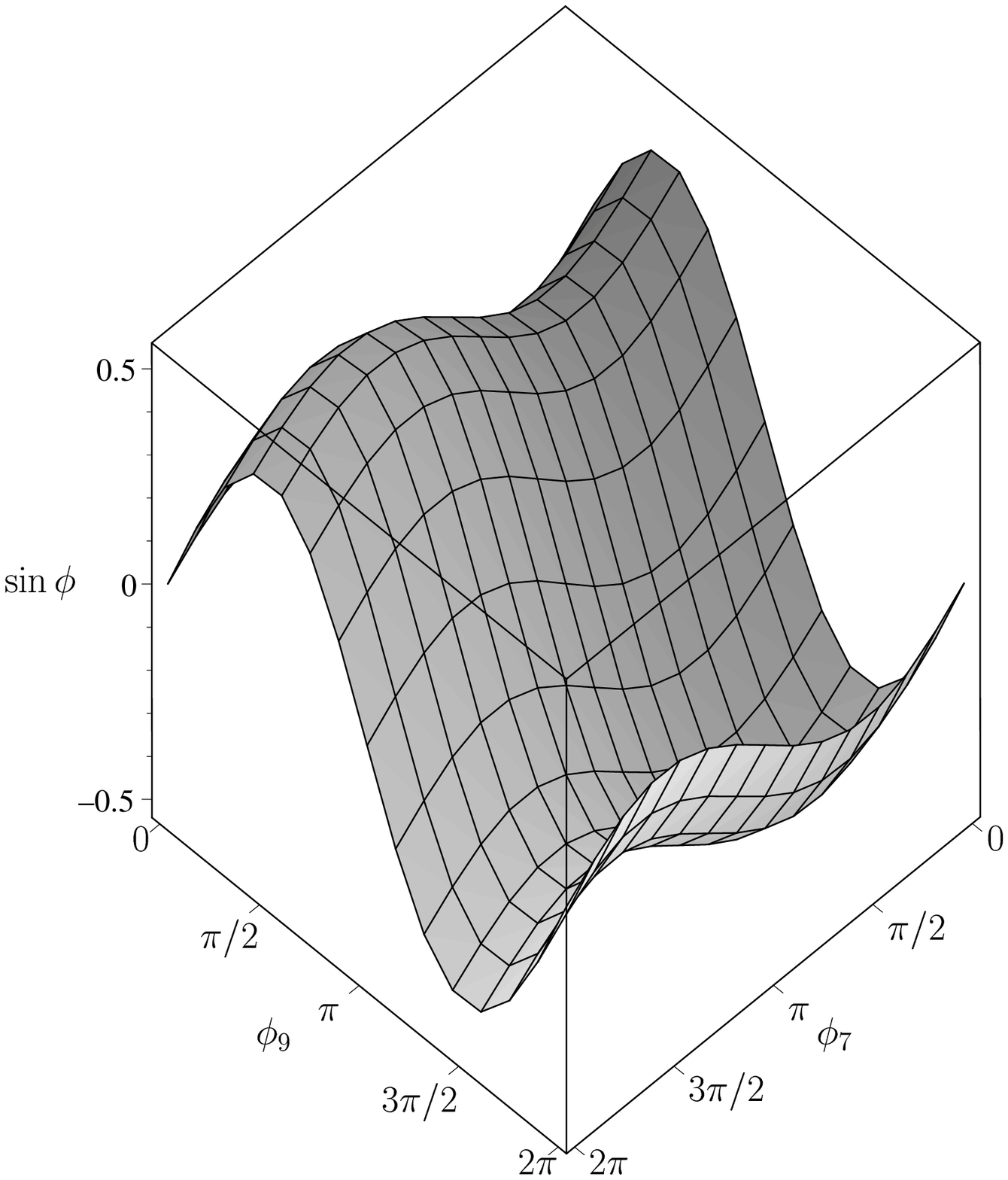,height=3.4in}\vspace{0.5em}
\caption{The parameters $r$ and $\sin\phi$ [see \eq{def:CPasym:gen}] 
vs $\phi_7$ and $\phi_9$ in the low dimuon invariant 
mass region. For simplicity,  we have taken $|R_7|=|R_9| = 1$ 
while $R_{10}$ is chosen to coincide with the experimental upper limit on 
the non-resonant branching ratio, i.e. $\branch^{\nr}=4.0\times 10^{-6}$.\label{l-3D-rphi}} 
\end{center}
\end{figure}

\begin{figure}[p]
\begin{center}
\epsfig{file=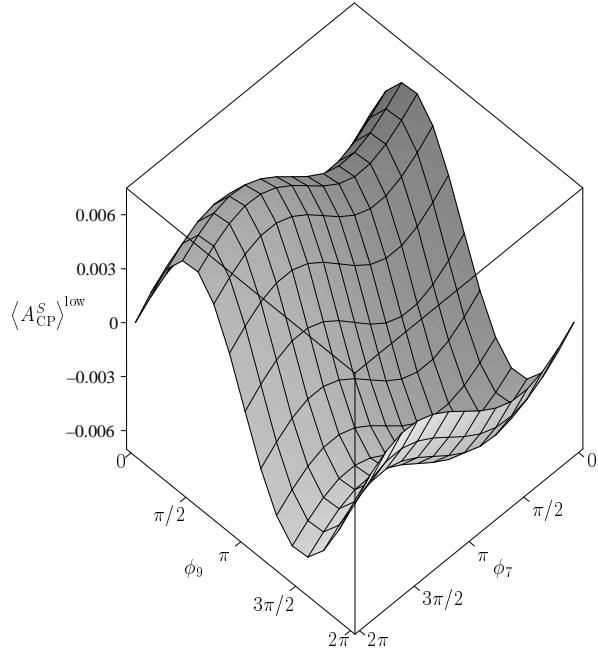,height=3.4in}\vspace{0.5em}
\caption{$\cp$-violating average asymmetry $\av{A_{\cp}^{S}}$ in the low 
dimuon invariant mass region vs $\phi_7$ and $\phi_9$, 
with the parameters $|R_i|$ ($i=7,9,10$)
as in \fig{l-3D-rphi}.\label{l-3D-ACP}}
\end{center}
\end{figure}

Figures \ref{l-3D-rphi} and \ref{l-3D-ACP} 
%
%
%
%
show the parameters $r$, $\sin \phi$, and $A_{\cp}^S$   
in the low dimuon invariant mass region as a function of the phases 
$\phi_7$ and $\phi_9$, taking $|R_9|=1$ and requiring that
$\branch^{\nr}=4.0\times 10^{-6}$.\footnote{The impact 
of new phases on the partial-rate asymmetry in $\B\to K^* l^+l^-$ decay 
for the case of $|R_i|=1$ has also been studied in \rf{aliev:etal}.}  
It may be noted that the predictions for the average $\cp$ asymmetry depend 
very little on the phase $\phi_7$ and its absolute value is no greater than
$1 \%$, regardless of the size of the $\cp$-violating phases.  
If we allow for deviations from $|R_9|=1$, the magnitude of the $\cp$ 
asymmetry does not change significantly. 
We may therefore conclude that the partial-rate asymmetry, indicative of 
$\cp$ violation, in the low-$\sh$ region is too small to be 
observable (assuming that the indispensable strong phase does not receive 
any substantial non-standard contributions). 

\begin{figure}
\begin{center}
\epsfig{file=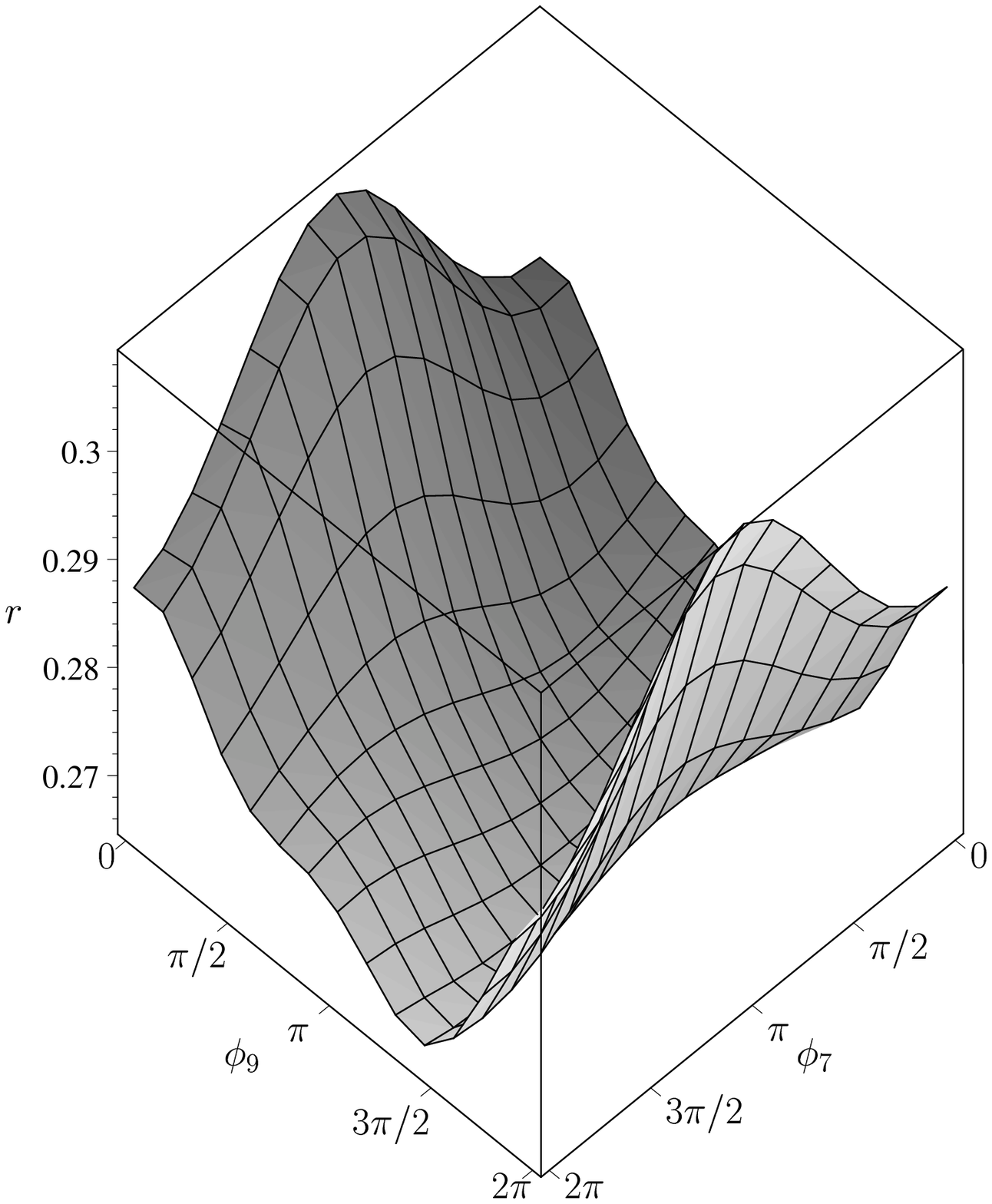,height=3.4in}\hspace{1em}
\epsfig{file=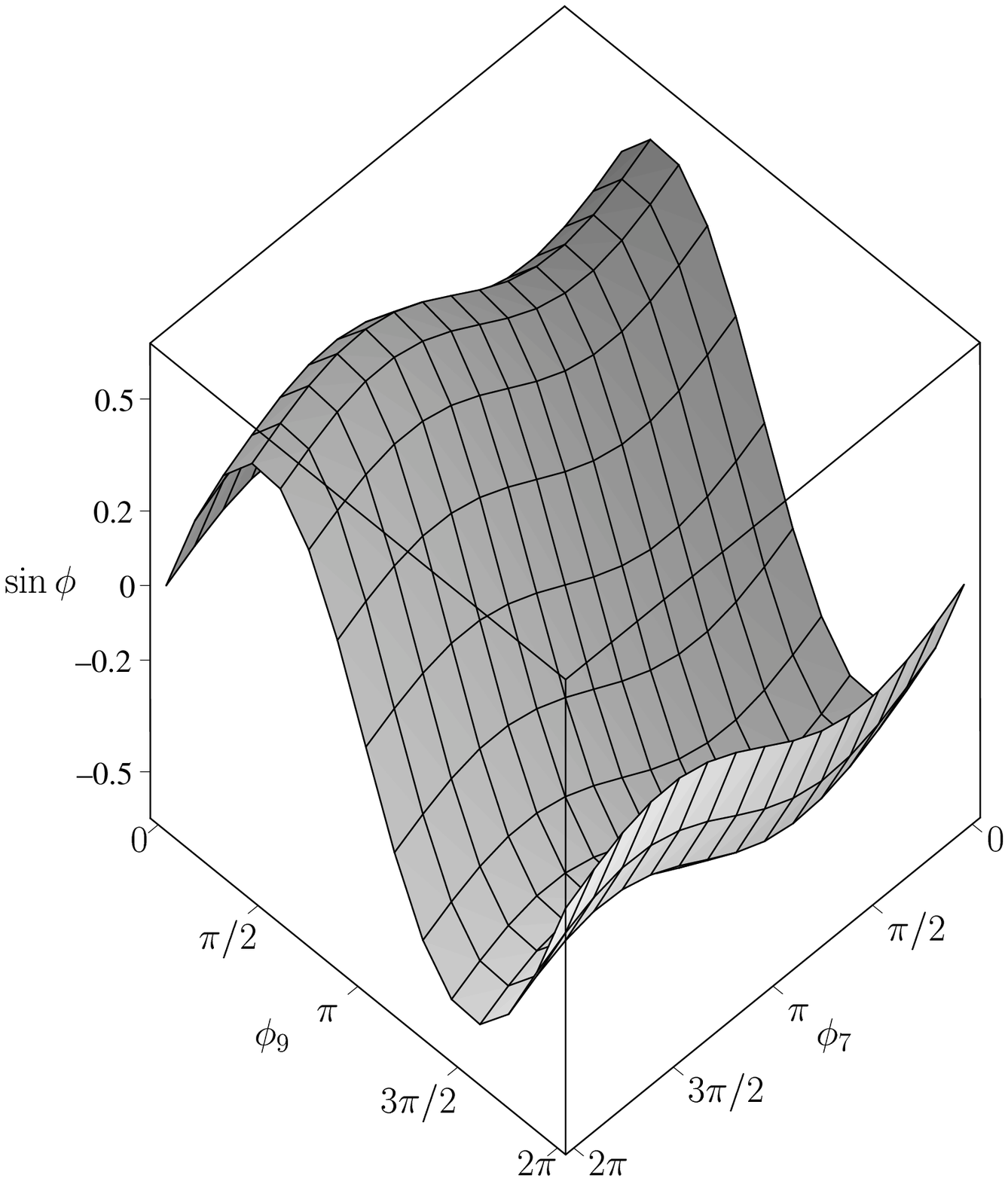,height=3.4in}\vspace{0.5em}
\caption{The parameters $r$ and $\sin\phi$ appearing in $A_{\cp}$, 
\eq{def:CPasym:gen}, as a function of $\phi_7$ and $\phi_9$ 
for the high dimuon invariant mass region, with $|R_7|=|R_9| = 1$.
The ratio $R_{10}$ is chosen to be consistent with the experimental 
upper limit on the non-resonant branching ratio given in 
\eq{cdf:limit}.\label{h-3D-rphi}}
\end{center}
\end{figure}

We now turn our attention to the high-$\sh$ region. 
It is clear from the above discussion
that physics beyond the SM can give rise to sizable $\cp$ asymmetries 
above the $\psi'$ resonance where we expect to have an appreciable 
strong phase but also a lower branching ratio. 
In fact, in the high-$\sh$ region  we find the approximate relation 
$r^{\high} \simeq 2 r^{\low}$, whereas the numerical value of the weak 
phase $\sin\phi$ is of the same order of magnitude in both the low and 
the high dimuon invariant mass region. 
This can be seen in Figs.~\ref{h-3D-rphi} 
%
%
%
%
\begin{figure}
\begin{center}
\epsfig{file=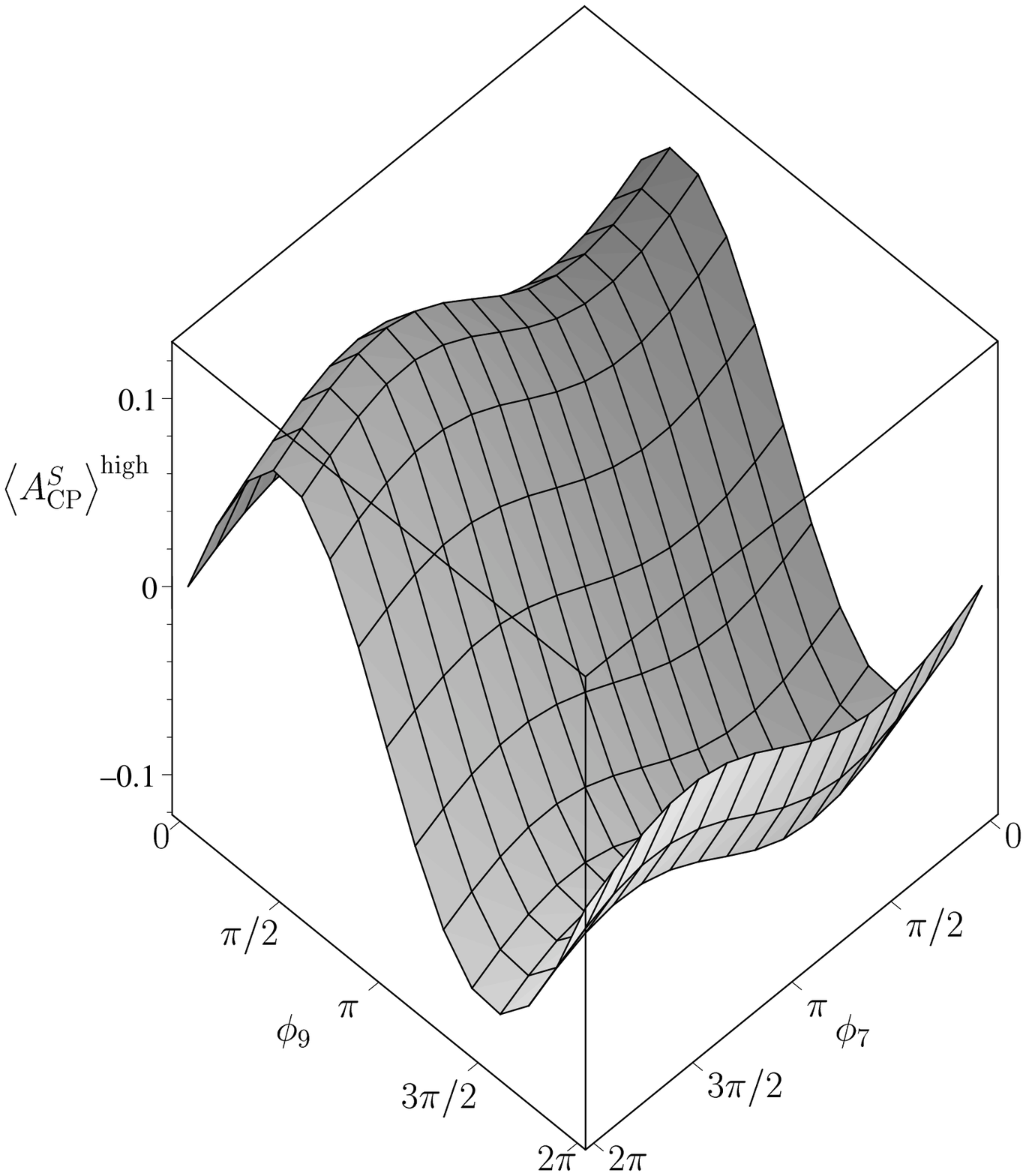,height=3.4in}\vspace{0.5em}
\caption{$\cp$-violating average asymmetry $\av{A_{\cp}^{S}}$ in the high 
invariant mass region of the muon pair, as a function of $\phi_7$ and 
$\phi_9$. The parameters $|R_i|$ ($i=7,9,10$) have been chosen as in 
\fig{h-3D-rphi}.\label{h-3D-ACP}}
\end{center}
\end{figure}
and \ref{h-3D-ACP}, where we show our results for
$r$, $\sin\phi$, and $A_{\cp}^S$. For certain values of the phases 
$\phi_7$ and $\phi_9$, 
the $\cp$ asymmetry can be of order $\pm 10 \%$. 
As far as theoretical uncertainties are concerned, we merely remark
that the numerical estimates for average $\cp$ asymmetries are largely 
independent of the parametrizations of form factors, so that  
the theoretical uncertainty associated with real $c\bar{c}$ 
intermediate states discussed in \Sec{effective:hamiltonian} gives by far 
the largest uncertainty in the predicted $\cp$ asymmetry. 

Next we consider the case where we allow for higher values of $|R_9|$.
We begin by noting that 
for $|R_9|> 1.75$ some part of the $(\phi_7, \phi_9)$ parameter space
is already excluded by the experimental 
upper bound on the non-resonant branching ratio, \eq{cdf:limit}. 
Exploiting the fact that $A_{\cp}^S$ depends only weakly on 
the phase $\phi_7$, we may take $\phi_7=4.8$ (see \fig{h-3D-rphi}).
Then, setting $|R_9|=1$ we find a magnitude of the average $\cp$ asymmetry 
varying from $-0.05$ to $0.12$, whereas for $|R_9| = 1.75$ we predict the 
range  $-0.15 \leqslant \av{A_{\cp}^S}\leqslant 0.20$. In this latter 
case, numerical values of $r$ between  
$0.26$ and $0.33$ are estimated while $\sin\phi$ can be 
maximal. In particular, for $|R_9|=1.75$ the weak phase $\sin\phi$ 
is nearly unity if we demand the non-resonant branching fraction to be 
$4.0\times 10^{-6}$.

Let us now discuss a scenario in which we abandon the assumption of 
having a non-resonant branching ratio of $4.0\times 10^{-6}$.
We focus here on the high-$\sh$ region as  
we do not expect any significant deviation from our results obtained
in the region below the charmonium resonances.
We first consider the case where the branching ratio is 
still fixed to the SM value of $\branch^{\nr}=1.8\times 10^{-6}$. 
Further, we take $\phi_7=4.8$ and 
$|R_9|=0.9$. (Note that larger values of $|R_9|$ are not consistent with 
the SM branching ratio).
As a result, the weak phase $\sin \phi$ exhibits almost the same 
$\phi_9$-dependent behaviour 
as in the previous case with maximum branching ratio presently 
allowed by experiment. In addition, a smaller value for $\branch^{\nr}$ 
leads to a wider range of $r$, 
namely $0.47 \leqslant r \leqslant 0.70$. 
As for $\cp$ violation, an average asymmetry of 
anything between $-0.20$ and $0.30$ is predicted.

It is also interesting to analyse the case in which the 
branching ratio is not fixed to any particular value
but is still compatible with the experimental results described above. 
Remembering that $R_{10}$ contributes only to the branching ratio, we set 
$R_{10}=0$ in order to get the highest possible value for the 
$\cp$ asymmetry (see \Sec{sec:cp:observables}). Consequently, 
the interference of the terms $R_7$ and $R_9$ now plays an essential  
role as the branching ratio diminishes for certain values of 
$\phi_9$, and so $r$ can be unity (which corresponds to the 
maximal size of the $\cp$ asymmetry). In this case, the $\cp$ asymmetry 
$\av{A_{\cp}^S}$ varies considerably and can take on any value between 
$-0.5$ and $0.4$ for $\branch^{\nr}$ ranging from 
$5.0\times 10^{-7}$ to $1.6\times 10^{-6}$.

\subsubsection{CP asymmetry in the angular distribution of $\m^-$}
\begin{figure}
\begin{center}
\epsfig{file=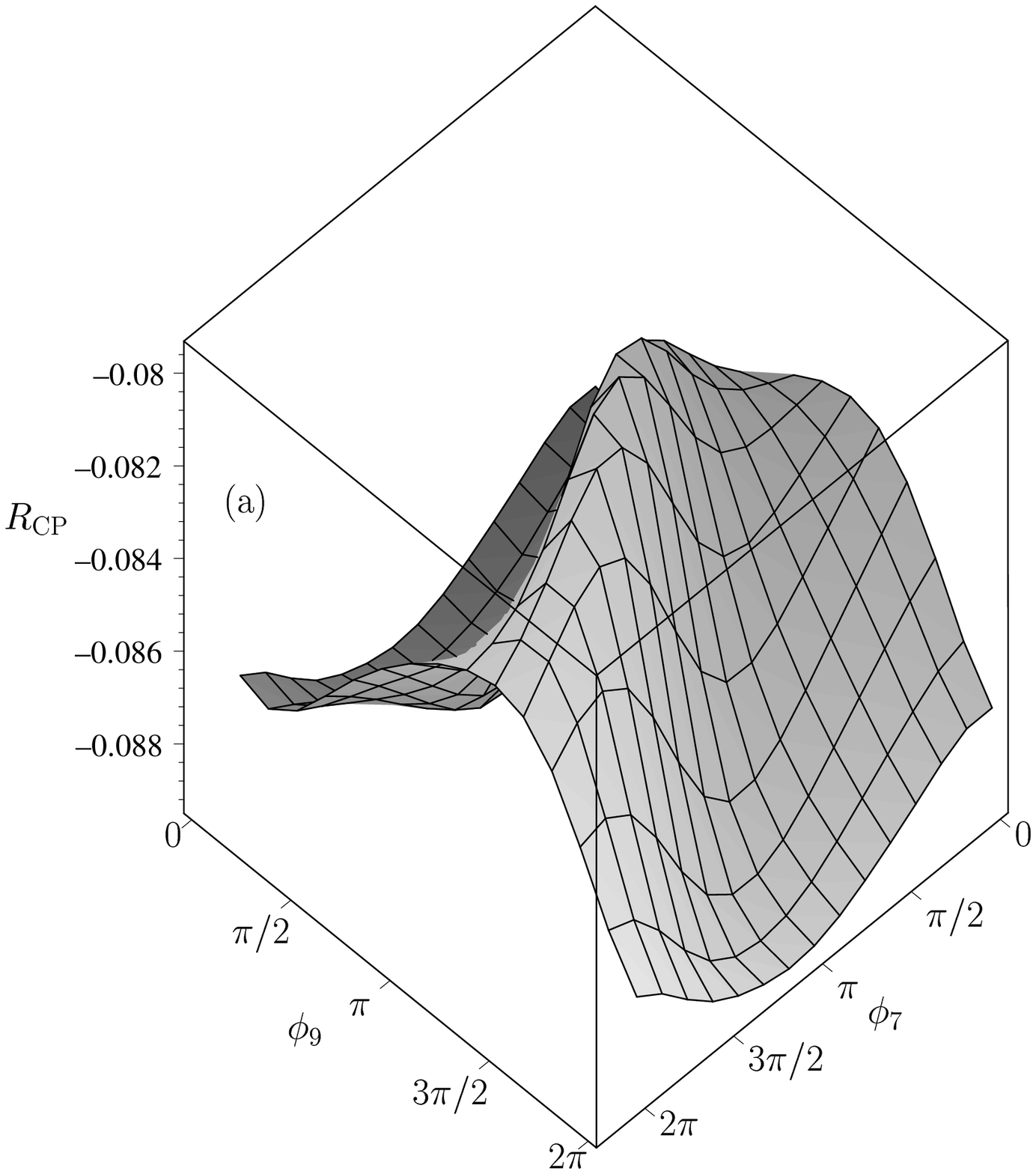,height=3.4in}\hspace{1em}
\epsfig{file=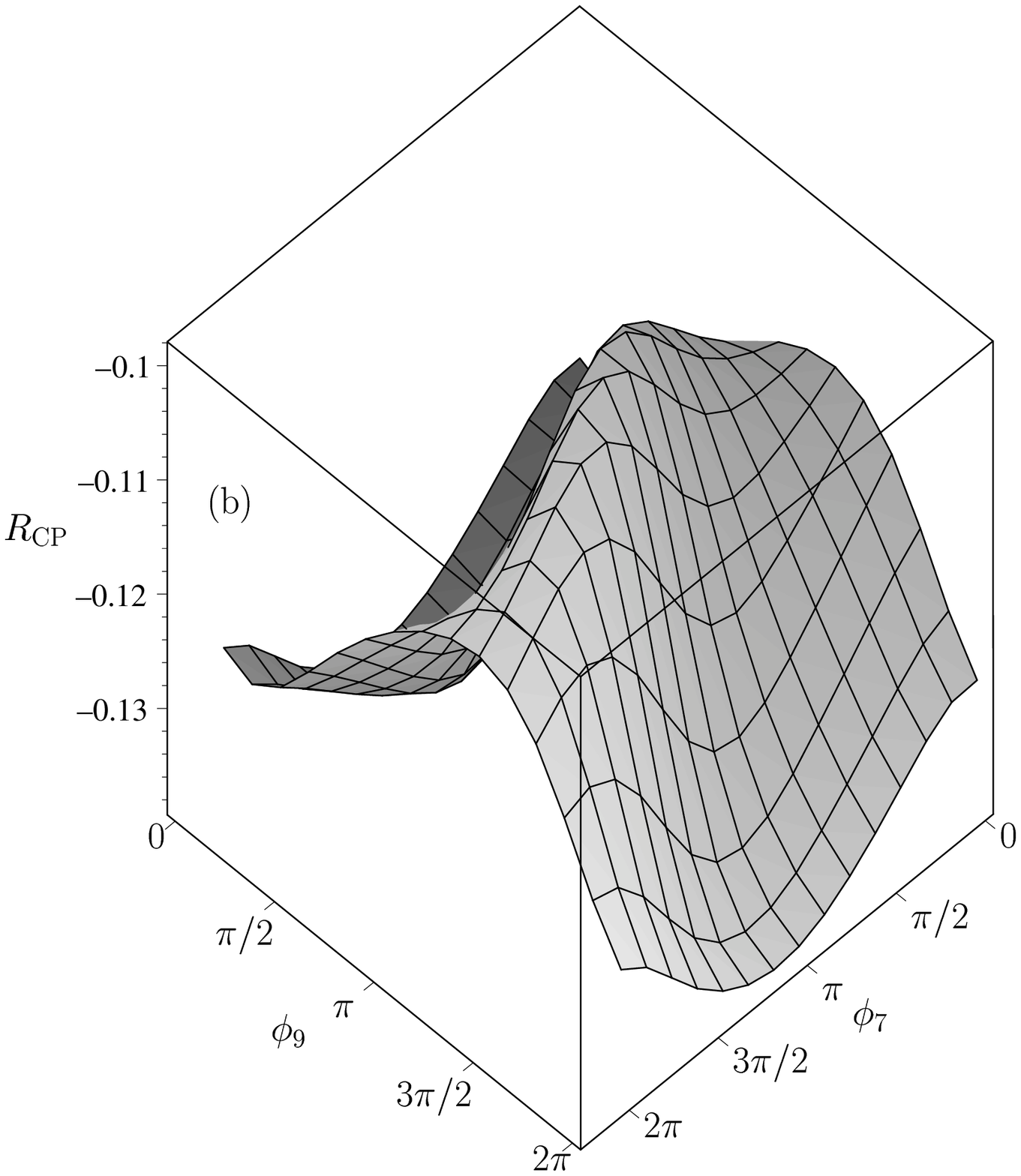,height=3.4in}\vspace{0.5em}
\caption{The quantity $R_{\cp}\equiv \av{A_{\cp}^{D}}^{\high}/\sin\phi_{10}$ 
[cf.~\eq{cp:asym:angular:high}] as a function of $\phi_7$ and 
$\phi_9$ with $|R_7|=1$.
(a) $|R_{10}|=1.9$ and $|R_9|$ is chosen such that it coincides with the 
upper limit on $\branch(B^0\to K^{*0} \m^+\m^-)$. (b) 
$|R_{10}|=1.2$ and the SM branching fraction of 
$1.8\times 10^{-6}$ has been adopted.\label{h-3D-FB}}
\end{center}
\end{figure}

A similar analysis has been carried out for the asymmetry 
$A_{\cp}^D$, \eq{def:CPasymSD}, which is of considerable interest 
since it probes the phase of $R_{10}$. Here again the $\cp$ asymmetry in the 
low-$\sh$ domain is fairly small, typically a few per cent, and
we therefore concentrate on the high-$\sh$ region where $\cp$-violating 
effects are not suppressed by small unitarity phases. Results for the ratio 
$R_{\cp}\equiv A_{\cp}^D/\sin\phi_{10}$ as a function of 
$\phi_7$ and $\phi_9$ are displayed in \fig{h-3D-FB}.
%
%
Demanding that $|R_9|$ reproduces the experimental upper bound on the 
non-resonant branching ratio leads to the predictions  shown in 
Figure \ref{h-3D-FB} (a). Moreover, the quantity $R_{10}$ obeys the 
constraint $|R_{10}| \leqslant  1.9$. On the other hand, assuming the 
SM prediction of $\branch^{\nr}=1.8\times 10^{-6}$, 
we obtain the results shown  in \fig{h-3D-FB} (b).
In this case, present experimental data on 
$\branch(B^0\to K^{*0} \m^+\m^-)$ lead to the upper bound 
$|R_{10}|\leqslant  1.2$. 
In both cases, the $\cp$ asymmetry in the angular distribution of $\m^-$ 
in $B$ and $\B$ decays turns out to be about $-10 \%$.\footnote{An analysis 
of such a $\cp$-violating effect in the presence of 
non-standard $Z$ couplings has recently been performed in 
\rf{buchalla:etal}, which estimates an asymmetry of about $10\%$ 
in the high dimuon  invariant mass region.}

Finally, taking $|R_9|=0$ we find that the $\cp$ asymmetry can be as 
large as $-25\%$ for a rather low $|R_{10}|$, 
which corresponds to a non-resonant 
branching fraction of $O(10^{-7})$. If we consider instead a branching 
ratio of $\branch^{\nr}\approx 10^{-6}$, 
the asymmetry can reach to values of $-15\%$.

\subsubsection{A comment on CP violation and additional operators}
As seen in the preceding, within the framework of the SM operator basis
it is possible to account for the maximum values of 
the $\cp$-violating asymmetries.
Hence our quantitative results for $\cp$ violation in the decay 
$\B\to K^* \m^+\m^-$ are not affected by the chirality-flipped
operators [\eq{operatorbasis:chirality}] once existing 
experimental constraints are taken into account. In other words, 
the observation of an appreciable $\cp$ asymmetry alone does not 
provide a test of the 
chirality structure of operators that enter the effective Hamiltonian, 
and thus is not sufficient to disentangle different new-physics scenarios.
%
%

\section{Summary and conclusions}\label{conclusions}
We have performed a largely model-independent analysis of the exclusive decay 
$\B\to K^* l^+l^-$ in the presence of physics transcending the SM.
In particular, we have investigated the implications of new $\cp$-violating 
phases for the decay $\B\to K^* \m^+\m^-$, and derived analytic expressions 
for the branching ratio, the FB asymmetry, and certain $\cp$-violating 
observables. The formalism presented is applicable to any effective 
Hamiltonian containing the SM operator basis
as well as operators with a different chirality structure.
We have studied in some detail the $\cp$ asymmetries
in the partial rates and the angular distribution of $\m^-$ in 
$\B\to K^* \m^+\m^-$ and $B\to \bar{K}^* \m^+\m^-$ decays, which 
require the simultaneous presence of weak and strong phases.
Adopting the SM operator basis and assuming that 
new-physics contributions to two-body non-leptonic $B$ decays are unlikely 
to be significant, the $\cp$-violating effects in the 
$2 m_{\m}\leqslant M_{\m^+\m^-}< M_{J/\psi}$ domain are estimated to be 
small (up to at most a few per cent). Ultimately, this result is a 
consequence of the smallness of the dynamical phase associated with the 
absorptive part of the penguin diagram; thus, the presence of large 
non-standard $\cp$-violating phases does not necessarily imply sizable  
$\cp$-violating effects in the lower part of the decay spectrum.
Even so, studies of $\cp$-violating effects in the low-$\sh$ region will 
provide a crucial test of the SM, since any indication of 
$\cp$ violation would represent new physics.
On the other hand, in the high dimuon invariant mass region 
$M_{\psi'}<M_{\m^+\m^-}\leqslant (M_B -M_{K^*})$ appreciable  
$\cp$-violating effects can show up, which are consistent with 
current experimental data on the $b\to s\g$ branching fraction and the upper 
bound on $\branch(B^0\to K^{*0} \m^+\m^-)$. We find that 
$\cp$ asymmetries up to $30\%$ can arise for a non-resonant branching 
ratio of $1.8\times 10^{-6}$. That is, new physics gives the same 
rate as in the SM while large $\cp$-violating effects may occur.
It should be kept in mind that our numerical results for the $\cp$ 
asymmetries in the high dimuon invariant mass 
region are plagued with theoretical uncertainties due mainly to 
real $c\bar{c}$ intermediate states, so that precise predictions are more difficult in this case.~Nevertheless, the $\cp$ asymmetries provide a particularly useful 
tool for discovering new physics
and their study may gain insight into the mechanism of $\cp$ violation.~Given 
an asymmetry of $20\%$ and a branching ratio of  $3.0\times 10^{-7}$ in the 
high-$\sh$ region, a measurement at $3\s$ level will 
necessitate at least $7.5\times 10^8$ 
$b\bar{b}$ pairs, which seems to be achievable in the $B$-factory era 
(see, e.g., \rf{b-exp}).
In this connection it is worth pointing out that the asymmetry 
$A_{\cp}^D$, which is a $\cp$-violating effect related to the angular 
distribution of $\m^-$ in $B$ and 
$\B$ decays, has the piquant feature that it does not require flavour 
identification and can be obtained from a measurement of the sum of 
$B$ and $\B$ events.
 
As far as new operators are concerned, we have argued that in the 
case of massless leptons, i.e.~$l=e$ or $\m$, the dominant contributions 
may come from operators with a non-SM chirality structure.
However, since the SM operator basis can accommodate maximum
$\cp$ asymmetries, 
the inclusion of new operator 
structures does not affect the main conclusions of our analysis,
but it is worth considering once sufficient data are accumulated.
We are thus eagerly awaiting the upcoming $B$ experiments which will 
provide useful information on the short-distance coefficients 
governing FCNC processes like $b\to s \g$ and $b\to s l^+ l^-$.   

We conclude that the $\cp$ asymmetries in the exclusive decay 
$\B\to K^* l^+ l^-$ can serve as an important test of the SM mechanism of 
$\cp$ violation and hence provide a testing ground for new physics.
A measurement of these asymmetries in forthcoming $B$ experiments 
would signal the presence of non-standard physics and rule out the SM as a 
primary source of $\cp$ violation.

\acknowledgments
One of us (F.K.) would like to thank the SISSA particle theory group for 
their warm hospitality during the early stages of this project. 
This work was supported in part by the TMR Network of the EC under 
contract ERBFMRX-CT96-0090. 
%
%

\end{document}